\documentclass{elsart}

  \usepackage{graphicx}
\newcommand{\beq}{\begin{equation}}
\newcommand{\eneq}{\end{equation}}

\begin{document}

\begin{frontmatter}



\title{Competing Boundary Interactions in a Josephson Junction Network
with an Impurity}


\author{Domenico Giuliano$^1$  and Pasquale Sodano$^2$}

\address{$^1$ Dipartimento di Fisica, Universit\`{a} della Calabria, Arcavacata di Rende
I-87036, Cosenza, Italy \\ and\\
INFN, Gruppo collegato di Cosenza, Arcavacata di Rende
I-87036, Cosenza, Italy\\
$^2$ Dipartimento di Fisica, Universit\`{a} di
Perugia, Via A. Pascoli, I-06123,
Perugia, Italy\\ and  \\ INFN, Sezione di Perugia, Via A. Pascoli, I-06123,
Perugia, Italy}

\begin{abstract}
We analyze a perturbation of the boundary Sine-Gordon model where
two boundary terms of different periodicities and scaling dimensions
are coupled to a Kondo-like spin degree of freedom. We show that, by
pertinently engineering the coupling with the spin degree of freedom,
a competition between the two boundary interactions may be induced,
and that this gives rise to nonpertubative phenomena, such as the emergence
of novel quantum phases: indeed, we demonstrate that the strongly coupled fixed point
may become unstable as a result of the ``deconfinement'' of a new
set of phase-slip operators -the short instantons- associated
with the less relevant boundary operator.
 We point out that a Josephson junction network with a pertinent
impurity located at its center provides a physical realization of
this boundary double Sine-Gordon model. For this Josephson junction network, we
prove that the competition between the two boundary interactions stabilizes
a robust finite coupling fixed point and, at a pertinent scale,
allows for the onset of $4e$ superconductivity.
\end{abstract}

\begin{keyword}
Boundary critical phenomena \sep Josephson junction arrays
\sep Quantum impurity models
 \PACS  05.30.Rt \sep 74.81.Fa \sep 74.50.+r
\end{keyword}
\end{frontmatter}

\section{Introduction}

There is a large number of physical systems that can be mapped
onto quantum impurity models in one dimension  \cite{boundary}.
Embedding a quantum impurity in a condensed matter system may
alter its responses to external perturbations \cite{hewson},
and/or induce the emergence of non Fermi liquid, strongly
correlated phases \cite{nonfl}. In quantum devices with tunable
parameters impurities may be realized by means of point contacts,
of constrictions, or by the crossing of quantum wires or Josephson
junction chains \cite{beenakker1,beenakker2,aoc,giuso4}.
While a standard perturbative approach works fine when  impurities
are weakly coupled to the other modes of the system (the
``environment''), there are situations in which  the impurities
are strongly coupled to the environment, affecting its behavior
through a change of  boundary conditions: when this happens,  it
is impossible to disentangle the impurity from the rest of the
system, the perturbative approach breaks down, and, consequently,
one has to resort to nonperturbative methods, to study the system
and the impurity as a whole. Such nonperturbative tools are
naturally provided by boundary  field theories (BFT)
\cite{boundary,jcardy}: BFTs  allow for deriving exact, nonperturbative
informations from simple, prototypical models which, in many
instances, provide an accurate description of experiments on
realistic low dimensional systems \cite{giamarchi}. In particular,
BFTs have been successfully used to describe  Josephson
current patterns in Josephson devices, such as chains with a weak
link \cite{glark,giuso1}, SQUIDs \cite{glhek,giuso2} and
$Y$ junctions \cite{giuso4}.

Motivated by the Kondo effect \cite{tscho},
impurity models have been largely studied to describe some magnetic chains
\cite{ager}, and static impurities in Tomonaga-Luttinger
liquids (TLL)s \cite{fencha}.  A renormalization group approach to those systems
leads, after bosonization \cite{gogolin}, to the investigation
of the phases accessible to pertinent boundary sine-Gordon models
\cite{fencha}. Scattering from an impurity often leads the boundary coupling
strength to scale to the strongly coupled fixed
point (SFP), which is rather simple since it describes a fully
screened spin in the Kondo system or a severed chain in the Kane-Fisher
model \cite{kanefish}. A remarkable exception is provided by
the fixed point attained in overscreened Kondo
problems, where an attractive finite coupling fixed point (FFP)
emerges in the phase diagram \cite{tscho}; this FFP is usually
characterized  by novel nontrivial universal indices and by
specific symmetries.
In the analysis of the Kondo effect,  an $SU(2)$ invariant coupling
of a local spin
degree of freedom with the spin density of conduction electrons,
 allows for engineering a marginally relevant interaction,
which would otherwise be irrelevant. Similar behaviors are
realized with crossed TLLs where, as a result of the crossing,
some operators turn from irrelevant to marginal, leading to
correlation functions exhibiting power-law decays with nonuniversal
exponents \cite{reyes,aoc}.

Superconducting Josephson devices  allow to engineer
remarkable realizations of the above situations,
\cite{giuso1,giuso2}. For superconducting
Josephson chains with an impurity in the middle  \cite{glark,giuso1}
or for SQUID devices \cite{glhek,giuso2} the phase diagram
admits only two fixed points: an unstable weakly
coupled fixed point (WFP), and a stable one at strong coupling,
while, for pertinent values of the fabrication
and control parameters, a FFP emerges
in  Y-shaped  Josephson junction  networks (JJN)s \cite{giuso4}.
The boundary field theory approach developed in
Ref.\cite{giuso1,giuso2} not only allows for an accurate
determination of the phases accessible to a superconducting
device, but also  for a field-theoretical treatment of the phase slips
(instantons),  describing quantum tunneling between degenerate
ground-states; furthermore,  it helps to evidence
remarkable analogies with models of quantum Brownian motion on
frustrated planar lattices \cite{kaneyi,saleurb}.

Here we study the effect of adding a less relevant scaling
operator to a boundary Sine-Gordon model. Most
analytical computations hold only when the second less relevant
operator has been scaled away \cite{cacardy}: conventional
wisdom suggests indeed that one should be able to neglect
all less relevant operators, when computing properties close
to the infrared fixed point. However, this expectation is
based only on weak coupling expansion, which can be quite misleading
\cite{saleurirr,azaria}. In this paper, we shall exhibit an explicit example
of a boundary field theory model where the added perturbation
may become relevant at strong coupling and we shall provide a superconducting
device where the onset of new nonperturbative phenomena may be observed.
Adding to a boundary Sine-Gordon model a perturbation with
a different scaling dimension and periodicity allows,
in a superconducting device, to change the tunneling charge and,
thus, to affect the transport across the device.
For quantum Hall fluids \cite{noisehalltheory}, superconductor-normal
metal contacts \cite{snortheo} and Kondo quantum dots \cite{sela},
adding a perturbation
modifies the charge of the excitations ,
as evidenced in dc shot noise
measurements \cite{noisehallexpe,nms,kesp}.

\begin{figure}
\includegraphics*[width=1.0\linewidth]{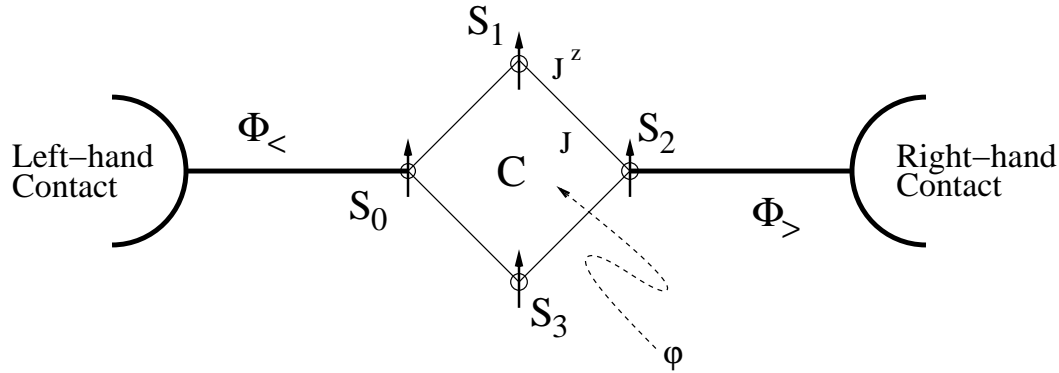}
\caption{The Josephson device: the central region {\bf C} with
two leads  connected to two external contacts.}
\label{device}
\end{figure}

We shall consider a boundary field theory with two boundary terms,
of different periodicities and scaling dimensions, coupled
to a Kondo-like spin degree of freedom. The resulting model is
described by a boundary double Sine-Gordon (BDSG) Hamiltonian,
given by $H_{\rm BDSG} =
H_{\rm LL} + H_{\bf B}$, where $H_{\rm LL} $ is a spinless
one-dimensional Tomonaga Luttinger Hamiltonian \cite{shulz}
-defined on a support of length $L$, with velocity $u$ and
Luttinger parameter $g$- given by

\beq
H_{\rm LL} = \frac{g}{ 4 \pi } \: \int_0^L \: d x \: \left[ \frac{1}{u}
\left( \frac{ \partial \Phi}{ \partial t } \right)^2 + u \left( \frac{
\partial \Phi}{ \partial x} \right)^2 \right]
\:\:\:\: , \label{el1} \eneq \noindent and \beq H_{\bf B} = - g_1
\: {\bf S}^z \: \cos [ \Phi ( 0 ) ] - g_2 \cos [ 2 \Phi ( 0 )] -
B_\parallel {\bf S}^z - B_\perp {\bf S}^x \:\:\:\: , \label{el2}
\eneq \noindent describing the  interaction between the Luttinger
field $\Phi$ and a spin-1/2 degree of freedom, localized at $x=0$.
In this paper,  we shall show that one can engineer the coupling
with the spin degree of freedom, so as to induce a competition
between the two periodicities in $H_{\bf B}$, leading, in some
instances, to the emergence of new quantum phases. We shall show
indeed that, for $1<g<4$ and for $B_\parallel = B_\perp = 0$, the
less relevant interaction $-g_2 \cos [ 2 \Phi ( 0 ) ]$
destabilizes the strongly coupled fixed point, as a result of the
``deconfinement'' of new phase-slip operators (instantons),
characteristic of the double Sine-Gordon interaction \cite{dsg2}.
To fix the ideas, we analyze in detail the Josephson junction
network depicted in Fig.\ref{device}, since it provides a
remarkable physical realization of the BDSG model described by
$H_{\bf B}$; in this JJN, we show that the competition between the
two periodicities in $H_{\bf B}$ stabilizes a robust
\cite{novais,giusonew} FFP, and -at a pertinent scale- allows for
the emergence of $4e$ superconductivity \cite{giusoepl}.

The paper is organized as follows:

In section \ref{model}, we show that the JJN in Fig.\ref{device}
is indeed described by $H_{\bf B}$, i.e., by a double boundary
Sine-Gordon Hamiltonian coupled to a pertinent spin-1/2 local spin
degree of freedom;

In section \ref{rengrou}, we  determine the phase diagram of the
DBSG model, using the renormalization group (RG) approach  and
show that it admits  a WFP, a strongly coupled fixed point (SFP),
and, for $1<g<4$ and for $B_\parallel = B_\perp = 0$ , a FFP.
Furthermore, we show that, near by the FFP, the emerging local
spin degree of freedom is robust against decoherence;

Section \ref{dcjoscur} is devoted to the analysis of Josephson current patterns
exhibited by the JJN. There we show that $4e$ superconducting correlations may
be probed in a Josephson current measurement,
in all the phases accessible to the JJN;

In section \ref{dctra1}, we evidence that a shot noise
measurement can account for the emergence of $4e$ tunneling
charges in the JJN, near by the WFP.  Furthermore, to show
that $4e$ superconductivity is a feature of the JJN also
far from the WFP, we derive an exact formula for the dc current, as
well as for the shot noise, at the ``magic point'' $g=2$ \cite{ameduri}, where
the WFP is not IR stable;

Section \ref{concl} is devoted to our concluding remarks, while
the appendices provide the necessary mathematical background for the
analysis carried in the paper.

\section{The boundary double Sine-Gordon Hamiltonian}
\label{model}

In this section, we show that the JJN depicted in
Fig.\ref{device}, may be effectively described by  $H_{\rm BDSG}$,
defined in Eqs.(\ref{el1},\ref{el2}). In Eq.(\ref{el2}), $g_1$ and
$g_2$  are real parameters, with $g_2 > 0$, while ${\bf S}^z $ and
${\bf S}^x$ are, respectively, the $z$-component and the
$x$-component of a spin-1/2 operator. $B_\parallel$ and $B_\perp$
may be regarded  as the two components -along $z$ and $x$,
respectively- of an external magnetic field acting on ${\bf
S}$; as such they may be regarded as control parameters to
tune the onset of different regimes.

The spin-1/2 degree of freedom allows for $H_{\rm BDSG}$ to be
invariant under

\beq
\tau_1 : \biggl\{ \begin{array}{c} \Phi  \longrightarrow  \Phi + \pi k , \\
                    S^z \longrightarrow S^z \: (-1)^k
                  \end{array}
\:\:\:\: , \label{ela.3} \eneq \noindent which realizes the usual
``Sine-Gordon symmetry'' with period $\pi$ (instead of $2\pi$);
for $k$ odd, $\tau_1$ involves also the sign inversion of ${\bf
S}^z$ \footnote{Notice that this is consistent with keeping ${\bf
S}^x$ unchanged, as one may change sign to two components of ${\bf
S}$, say ${\bf S}^z$, ${\bf S}^y$ (not appearing in $H_{\bf B}$),
without altering the canonical commutation relations}. As we shall
see, the emergence of this symmetry is crucial to account for the
novel behaviors in the JJN depicted in Fig.\ref{device}.

The JJN  consists of a central rhombus {\bf C}, made with
four Josephson junctions of nominal strength $J$, pierced by a
dimensionless flux $\varphi$ (i.e., $\varphi = \Phi / \Phi_0^*$)
and connected to two chains (leads) of  Josephson junctions, of nominal strength $E_J$,
with charging energy $E_c \gg E_J$, and charge repulsion
strength between nearest-neighboring junctions given by $E^z$.
The gate voltage applied to each junction
is tuned at the degeneracy between charge eigenstates with
$N$ and $N+1$ Cooper pairs, so that each junction may be regarded as
an effective spin-1/2 variable. In this
regime, the low-energy, long wavelength dynamics of the two leads
is well described in terms of two LL Hamiltonians for the
plasmon fields of the chain on the left- and the right-hand side
respectively, $\Phi_< , \Phi_>$;  the Luttinger parameters $g$ and $u$ are
given by $g = \frac{\pi}{2 ( \pi - {\rm arccos}
( \frac{\Delta}{2} ))}$,  $u = v_f \left[ \frac{\pi}{2} \frac{\sqrt{1 - (\frac{\Delta}{2})^2}}{
{\rm arccos ( \frac{\Delta}{2} )}
} \right]$ ($\Delta = (E^z - 3 E_J^2 / 16 E_c) / E_J$) \cite{glark,giuso1}.
The central region {\bf C} is described by
$H_{\bf C} = - J  \sum_{ j = 0}^3 \{ e^{ i \frac{\varphi}{4} } S_j^+
S_{j+1}^- + {\rm h.c.} \} + J^z  \sum_{ j = 0}^3  S_j^z
S_{j+1}^z$, with $\vec{S}_j$ spin-1/2 variables defined at
site $j$ . To trade $H_{\bf C}$ for an effective boundary
interaction, one performs a systematic Schrieffer-Wolff (SW)
sum over the high-energy eigenstates of $H_{\bf C}$.
This is carried out in appendix \ref{swolff}
where it is shown that, for $\varphi = \pi$,
the ground state of $H_{\bf C}$
is twofold degenerate, with the two degenerate states
given by $ | \Uparrow \rangle =
\frac{1}{2 \sqrt{2}} \{  \sqrt{2} [  | \uparrow \downarrow
\uparrow \downarrow \rangle +  | \downarrow \uparrow \downarrow \uparrow
\rangle ] +  | \downarrow \downarrow \uparrow \uparrow \rangle
+   | \uparrow \uparrow \downarrow \downarrow \rangle +
 | \uparrow \downarrow \downarrow \uparrow \rangle +
 | \downarrow \uparrow \uparrow \downarrow \rangle
 \}$, and by $| \Downarrow \rangle  = \frac{1}{2 \sqrt{2}} \{ \sqrt{2} [ | \uparrow \downarrow
\uparrow \downarrow \rangle - | \downarrow \uparrow \downarrow \uparrow
\rangle ] - i | \downarrow \downarrow \uparrow \uparrow \rangle
- i  | \uparrow \uparrow \downarrow \downarrow \rangle +
i | \uparrow \downarrow \downarrow \uparrow \rangle +
i | \downarrow \uparrow \uparrow \downarrow \rangle
 \}$.

The degeneracy between $  | \Uparrow \rangle $ and
$ | \Downarrow \rangle $ may be  removed by slightly detuning $\varphi$
to  $\varphi = \pi + \delta$, with $| \delta / \pi | \ll 1$.
From Eq.(\ref{metel5.a}), one sees that removing
the degeneracy induces  the term $- B_\parallel
{\bf S}^z$ in $H_{\bf B}$; at variance, detuning the gate
voltage applied to the junctions in {\bf C}  yields the term  $- B_\perp
{\bf S}^x$ (see Eq.(\ref{metel5.b})).

Connecting the leads to
{\bf C} with two Josephson junctions
of nominal strength $\lambda$ ($\ll E_J , J$), allows
-via the SW procedure described  in appendix \ref{swolff}-
to determine the effective boundary Hamiltonian, which, to
the fourth order in $\lambda$, coincides with
Eq.(\ref{el2}), with $B_\parallel =
- 4 J \sin \left( \frac{\varphi - \pi}{4} \right) $,
$B_\perp = \frac{  \sqrt{2} ( 2 + \sqrt{2} ) \lambda^2}{ ( 2 - \sqrt{2} )^2 }
\frac{ \lambda^2 h}{J^2} $,
$\Phi = ( \Phi_< - \Phi_> ) / \sqrt{2}$,
$g_1 = \frac{\lambda^2}{2 J }  \left(
\frac{1 + \sqrt{2}}{2 - \sqrt{2}} \right) $, $g_2 =
2 \frac{ C \lambda^4}{  J^3} $, $C$ being a numerical
coefficient $\sim 10^{-1}$.
As evindenced  in Ref.\cite{giusoepl},
the  first term in  $H_{\bf B}$  describes tunneling of Cooper pairs
between the two leads of the device,
while the second term is responsible for the
coherent tunneling of pairs of Cooper pairs
across {\bf C}.

While $H_{\bf B}$ provides the dynamical boundary
conditions (BC)s at the inner
boundary, the BCs at the outer boundary ($x=L$) depend on the
type of external contacts one attaches to the JJN to
induce a current across the leads.
In particular, when the JJN is connected
to two metallic leads at a finite voltage bias $V$
(or not contacted), one may safely assume Neumann BCs
($\frac{\partial \Phi ( L )}{ \partial x} = 0$)
at the outer boundary while, when the
device is connected to two bulk superconductors at fixed
phase difference $\alpha$ (as it happens when a
dc Josephson current is induced across {\bf C}), one may assume
Dirichlet-like BCs ($\Phi ( L , t ) = \alpha$)
at the outer boundary.

\section{Perturbative renormalization group analysis}
\label{rengrou}

In this section, we  use the RG approach to investigate the phase
diagram accessible to a system described by  $H_{\rm DBSG}$. A
perturbative analysis of  the boundary interaction shows that
there is a range of values of $g$ for which $H_{\bf B}$ becomes a
relevant operator and, furthermore, evidences the effects of its
two competing harmonics. We shall show that the phase diagram
admits a WFP, an SFP, and, for $1<g<4$ and for $B_\parallel =
B_\perp = 0$, it allows for the emergence of a FFP, which is
responsible for some of the remarkable novel behaviors exhibited
by the JJN in Fig.\ref{device}.

In subsection \ref{rg1}, we derive the perturbative RG equations
at weak coupling and use them to investigate the stability of the
WFP; in subsection \ref{rg2}, we repeat the same analysis near by
the SFP and, in subsection \ref{rg3}, we show that a stable finite
coupling fixed point emerges within a pertinent window of values
of $g$ when $B_\parallel = B_\perp = 0$; finally, we
investigate how decoherence may be frustrated \cite{novais} when
the JJN is operated near by the FFP.

\subsection{Perturbative renormalization group analysis near by the WFP}
\label{rg1}

To derive the perturbative RG equations near by the WFP requires computing
the partition function of the system by
integrating over the field $\Phi$, as well as over the spin
variable ${\bf S}$. To perform the latter integration,
we resort to the imaginary time formalism and introduce
two local complex fermion variables, $\{ a , b \}$,
to describe the spin-1/2 operator, which is then
given by  ${\bf S}^z \longrightarrow a^\dagger a - b_\dagger b$, and
${\bf S}^x \longrightarrow a^\dagger b + b^\dagger a$. As a
result, the  Euclidean action for the boundary degrees of freedom of the system
is given by

\beq
S_{\bf B} = S_{\bf B}^{(0)} + S_{\bf B}^{(I)}
\:\:\:\: ,
\label{compl0.1}
\eneq
\noindent
where
\beq
S_{\bf B}^{(0)}   = \int_0^\beta \: d \tau \: \left\{ a^\dagger \left[ \frac{\partial }{ \partial \tau}
- i \omega_0 - B_\parallel\right] a + b^\dagger \left[ \frac{\partial }{ \partial \tau}
- i \omega_0 + B_\parallel\right] b \right\}
- B_\perp \int_0^\beta \: d \tau \: \{ a^\dagger b + b^\dagger a \}
\:\:\:\: ,
\label{compl0.2}
\eneq
\noindent
and
\beq
 S_{\bf B}^{(I)} =
- g_1 \int_0^\beta \: d \tau \: \{ a^\dagger a - b^\dagger b \} \cos [ \Phi ( \tau ) ]
- g_2 \int_0^\beta \: d \tau \: \cos [ 2 \Phi ( \tau ) ]
\:\:\:\: ,
\label{compl1}
\eneq
\noindent
with  $\omega_0 = \pi / \beta$, $\beta = (k_B T)^{-1}$, and  $\Phi ( \tau ) = \Phi ( 0 , t = i \tau ) $
\cite{novais}.  The partition function is then given by

\beq
{\bf Z} = {\bf Z}_0 \: \langle {\bf T}_\tau e^{ - S_{\bf B}^{(I)} } ] \rangle_{(0)}
\:\:\:\: ,
\label{compl2}
\eneq
\noindent
with   ${\bf Z}_0 = {\rm Tr} \exp [ - \beta ( H_{\rm LL} - B_\parallel {\bf S}^z -
 B_\perp {\bf S}^x)]$,  ${\bf T}_\tau$ being the imaginary time-ordering product
operator, and $\langle \ldots \rangle $ denotes thermal averaging with
weight function $\exp [ - \beta ( H_{\rm LL} - B_\parallel {\bf S}^z -
 B_\perp {\bf S}^x)] / {\bf Z}_0$.

To integrate over the local fermion operators, one
needs to determine the relevant imaginary time correlation
functions of ${\bf S}^x , {\bf S}^z$;
these are given by

\beq
\langle {\bf S}^z \rangle_{(0)}  = \cos ( \theta ) \;\;\; , \;\;
\langle {\bf S}^x \rangle_{(0)}  = \sin ( \theta )
\:\:\:\: ,
\label{ccompl1}
\eneq
\noindent
with $\cos ( \theta ) = B_\parallel / \sqrt{B_\parallel^2 +
B_\perp^2}$, $\sin ( \theta ) = B_\perp / \sqrt{B_\parallel^2 +
B_\perp^2}$, and

\begin{eqnarray}
\langle {\bf T}_\tau [ {\bf S}^z ( \tau ) {\bf S}^z ( \tau^{'} ) ] \rangle_{(0)} &=&
\cos^2 ( \theta ) + \sin^2 ( \theta ) e^{ - 2 \lambda | \tau - \tau^{'} | }
\nonumber \\
\langle {\bf T}_\tau [ {\bf S}^x ( \tau ) {\bf S}^x ( \tau^{'} ) ] \rangle _{(0)} &=&
\sin^2 ( \theta ) + \cos^2 ( \theta ) e^{ - 2 \lambda | \tau - \tau^{'} | }
\;\;\;\; ,
\label{ccompl2}
\end{eqnarray}
\noindent with $\lambda = \sqrt{ B_\parallel^2 + B_\perp^2} $.
From Eqs.(\ref{ccompl1}), one sees that  $B_\parallel = 0
\; (B_\perp =0) \Rightarrow \cos ( \theta ) = 0 \; (\sin ( \theta) =0)
\Rightarrow \langle {\bf S}^z \rangle_{(0)} = 0 \: ( \langle {\bf S}^x
\rangle_{(0)}=0)$.

To integrate  over $\Phi$, one has to specify its BCs at
both boundaries. At $x=0$, the pertinent BCs are set  by
energy conservation, which amounts to require

\beq
\frac{u \pi}{2 \pi} \frac{\partial \Phi ( \tau )}{\partial x} = g_1 {\bf S}_G^z \sin [ \Phi ( \tau ) ]
+ 2 g_2 \sin [ 2 \Phi ( \tau ) ]
\:\:\:\: .
\label{bcon1}
\eneq
\noindent
Within a perturbative approach in
$H_{\bf B}$, one should then assume Neumann BCs
at $x=0$ (i.e., $\frac{\partial \Phi ( \tau )}{\partial x} =0$),
and require free BCs at $x=L$ (i.e., $\frac{\partial \Phi ( L , \tau )}{\partial x} =0$).

Neumann BCs at both boundaries yield  the following
mode expansion for  $\Phi ( x , \tau )$

\beq
\Phi ( x , \tau ) = \frac{1}{\sqrt{g}} \left[ \phi_0 + \frac{ 2 \pi i u \tau}{L} \tilde{P} \right]
+ i \sqrt{\frac{2}{g}}
\left\{ \sum_{ n \neq 0} \frac{a_n}{n} \: \cos \left[ \frac{2 \pi n x}{L} \right] e^{ - \frac{2 \pi n u \tau }{L}}
\right\}
\:\:\:\: ,
\label{el5}
\eneq
\noindent
with $ [ \phi_0 , \tilde{P} ] = i$, and $ [ \alpha_n ,
\alpha_m ] = n \delta_{n+m , 0}$.
By substituting Eq.(\ref{el5}) into Eq.(\ref{compl1}),
and normal-ordering the vertex operators  with respect to the ground state of
$H_{\rm LL}$, one gets

\beq
S_{\bf B} ( \tau ) = - \int_0^\beta \: d \tau \:
\{ \bar{g}_1 [ a^\dagger ( \tau ) a ( \tau ) -
b^\dagger ( \tau ) b ( \tau ) ] : \cos [ \Phi ( \tau ) ] :  +
\bar{g}_2 : \cos [ 2 \Phi ( \tau )] :  \}
\:\:\:\: ,
\label{el4}
\eneq
\noindent
with $\bar{g}_{1} = \left( \frac{a}{L} \right)^{\frac{1}{g}} g_1$, and
$\bar{g}_{2} = \left( \frac{a}{L} \right)^{\frac{4}{g}} g_2$,
$a / u$ being a pertinent short (imaginary time)
distance cutoff.

From Eq.(\ref{el4}), using the  standard factorization formula
of vertex operators \cite{pginsp}, one gets that

\beq
\langle {\bf T}_\tau [ : e^{ i \alpha_1 \Phi ( \tau_1 ) } :  \ldots \: : e^{ i \alpha_n \Phi ( \tau_N ) } : ] \rangle_0
= \exp \left[ \sum_{ i < j =1}^n \:2  \frac{\alpha_i \alpha_j}{g} \: \gamma_\tau ( \tau_i , \tau_j ) \right]
\: \delta_{ \sum_{ i  =1}^n \alpha_i , 0 }
\:\:\:\: ,
\label{el7}
\eneq
\noindent
with

\beq
\gamma_\tau ( \tau , \tau^{'} ) = \ln \left| 2 \sinh \frac{\pi u}{L} ( \tau - \tau^{'} )  \right|
\:\:\:\: .
\label{el8}
\eneq
\noindent
Eq.(\ref{el2}), implies that, in computing
${\bf Z}$ as a power series in $S_{\bf B}^{(I)}$,
vertex operators $:e^{ \pm i \Phi ( \tau )}:$ should
be always accompained  by an operator ${\bf S}^z ( \tau )$.
As a result, as long as $\cos ( \theta ) \neq 0$ and $\lambda \neq 0$,
and for length scales $L \geq L_\lambda  \sim \frac{\pi u}{\lambda}$,
one finds that

\beq
 {\bf Z} = {\bf Z}_0 \: \sum_{ n = 0}^\infty \: \frac{1}{n !} \: \prod_{ j = 1}^n [ \int_0^\beta \: d \tau_j ] \:
\sum_{ \alpha_j = \{ \pm 1 , \pm 2 \}}
\prod_{ \ell = 1}^n \left( \frac { y_{ | \alpha_\ell | } }{2} \right)
\: \langle {\bf T}_\tau [ : e^{ i \alpha_1 \Phi ( \tau_1 ) } : \ldots \: : e^{ i \alpha_n \Phi ( \tau_n ) } : ] \rangle_{(0)}
\:\:\:\: ,
\label{el6}
\eneq
\noindent
with

\beq
y_{ 1} = \cos ( \theta ) \bar{g}_1 \:\:\: , \:\:
y_{2} = \bar{g}_2 + \sin^2
 ( \theta ) \frac{ \Gamma [ 1 + 2/g]}{2 \lambda}
\left( \frac{\pi u }{4 a \lambda} \right)^{\frac{2}{g}}
\left( \frac{a}{L} \right)^{\frac{4}{g}} g_1^2
\:\:\:\: .
\label{pippo}
\eneq
\noindent
Eq.(\ref{pippo}) allows to infer the RG flow of the
boundary interaction, when $( \cos ( \theta)  ,
\sin ( \theta)  ) \neq ( 0 , 0 )$.
It should be noticed that ${\bf Z}$ in Eq.(\ref{el6}) may be
regarded as the partition  function of a one-dimensional
Coulonb and that the fugacities $y_1 , y_2$ associated
to its ``charges''  scale as $L^{- \frac{1}{g}}$ and as $L^{- \frac{4}{g}}$,
respectively.

\begin{figure}
\includegraphics*[width=.7\linewidth]{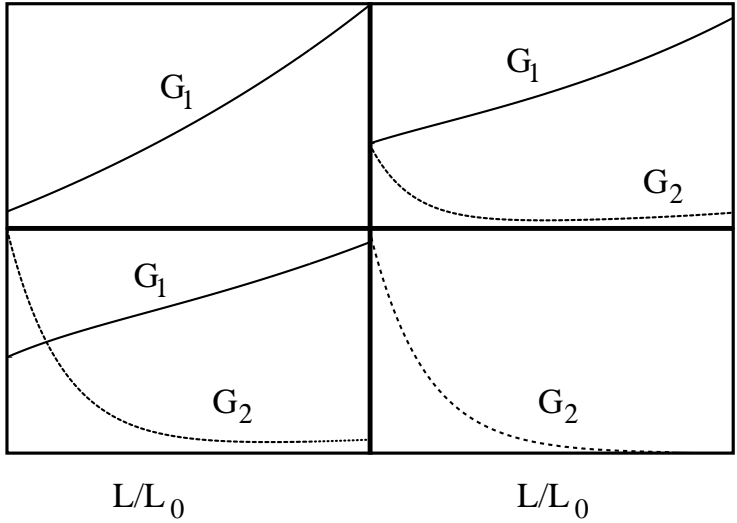}
\caption{Renormalization group trajectories of $G_1 ( L )$ and
$G_2 ( L)$, for $1<g<4$ and for $B_\parallel = B_\perp = 0$,
for different values of $G_1 ( L_0 )$ and
$G_2 ( L_0)$. {\bf Top left panel:} $G_1 ( L_0 ) \neq 0 , G_2 (L_0 )=0$;
{\bf Top right panel:} $G_1 (L_0 ) = G_2 (L_0 )$;
{\bf Bottom left panel:} $ G_1 (L_0 ) <G_2 (L_0 )$;
{\bf Bottom right panel:} $ G_1 (L_0 ) =0 , G_2 (L_0 ) \neq 0$.}
\label{rgflow}
\end{figure}

To study the behavior of the boundary interactions
along the RG trajectories, one needs to write down the
RG equations for the running coupling
strengths $G_1 (L) = \bar{g}_1 \: \left( \frac{L}{L_0} \right)$,
$G_2 ( L ) = \bar{g}_2 \left( \frac{L}{L_0} \right)$,
where $L_0$ is a reference length;  these equations
may be derived from the short (imaginary time) distance
operator product expansions
(O.P.E.)s of the vertex operators entering $H_{\bf B}$ \cite{jcardy},
which are given by

\begin{eqnarray}
\{ : e^{ \pm i \Phi ( \tau ) } : \: : e^{ \pm i \Phi ( \tau^{'} )} : \} & \approx_{ \tau^{'} \to \tau^- } &
\: \left[ \frac{  \pi u ( \tau - \tau^{'} )}{L} \right]^{\frac{2}{g} }: e^{ \pm 2 i \Phi ( \tau ) } :
+ \ldots \nonumber \\
\{ : e^{ \pm i \Phi ( \tau ) } : \: : e^{ \mp 2 i \Phi ( \tau^{'} )} : \} & \approx_{ \tau^{'} \to \tau^- }&
\: \left[ \frac{  \pi u ( \tau - \tau^{'} )}{L} \right]^{- \frac{8}{g} }: e^{ \mp i \Phi ( \tau ) } :
+ \ldots
\:\:\:\: ,
\label{el12}
\end{eqnarray}
\noindent
where the ellipses denote  non diverging contributions.
As a result, one gets

\begin{eqnarray}
\frac{ d G_1 }{ d \ln \left( \frac{L}{L_0} \right)} &=& \left[ 1 - \frac{1}{g}
\right] G_1 + G_1 G_2
\equiv \beta_1 [ G_1 , G_2]
\nonumber \\
\frac{ d G_2 }{ d \ln \left( \frac{L}{L_0} \right)} &=& \left[ 1 - \frac{4}{g}
\right] G_2 + \frac{1}{2} \frac{(G_1 )^2}{2}
= \beta_2 [ G_1 , G_2 ]
\:\:\:\: .
\label{el13}
\end{eqnarray}
\noindent
From Eq.(\ref{el13}), one sees that, for $g<1$ both $\beta_1$ and $\beta_2$ are
negative for small values of $G_1 , G_2$; thus, for $g<1$, $H_{\bf B}$ provides
an irrelevant boundary perturbation, and a perturbative analysis in the boundary
coupling is expected to yield consistent results. On the contrary, for
$g>1$, $\beta_1$ starts positive, and, despite the fact that
$\beta_2$ starts negative form small values of $G_1 ( L_0 ) , G_2 ( L_0 )$,
$\beta_2$ becomes positive at some
intermediate scale $L_{12} \approx \left[ ( \frac{4}{g} - 1 )
\frac{\bar{g}_2}{\bar{g}_1}
\right]^{\frac{g}{g-1}}$, since $G_1$ increases
as $L / L_0$ increases. Thus, $G_1$ renormalizes $G_2$,
so as to  make, even for $g<4$,
$\cos [ 2 \Phi ( 0 )]$ a relevant perturbation, despite the fact that  the linear
term of the $\beta_2$ function is negative.

Upon defining the  ``healing length'' $L_*$ by $G_i ( L_* ) \sim
1$ \footnote{Estimating $L_*$ requires, in general, to resort to a
numerical calculation. For a simple choice of the bare value of
the couplings, such as $G_1 (L_0) \sim G_2 ( L_0 ) \sim 2 y$, one
obtains $L_* \approx \frac{1}{1 - (1 / g) } \: \ln \left[ \frac{1
- ( 1 /g )  + y }{( 2 - ( 1 / g ) )y}\right]$.}, one finds that,
for $B_\parallel = B_\perp = 0$, and for $g>1$ and $L > L_*$, the
system is driven out of the perturbative regime, still preserving
its fundamental periodicity $\Phi \to \Phi + \pi$ (i.e.,
$\tau_1$-symmetry). As we shall see,  $\tau_1$-symmetry 
allows to classify the symmetries of the various infrared (IR)
stable fixed points accessible to the JJN.

\subsection{Renormalization group analysis near by the SFP}
\label{rg2}

In the previous subsection, we showed that, for $g>1$, the WFP becomes IR unstable and,
that, for $L \geq L_*$, the system is driven out of the (weakly coupled) perturbative
regime. To infer the IR properties, one may
safely assume that the boundary interaction is
driven all the way down to the SFP and, then, exploit standard
 boundary conformal field theory (BCFT) techniques \cite{aoc},
to build the leading boundary perturbation  at the SFP,
in order  to derive the pertinent RG equations.

When $B_\parallel \neq 0$ and $1<g<4$, the running coupling $G_1 ( L)$
grows with $L$, while $G_2 $ is an irrelevant coupling.
As a result, for $L \geq L_*$, the system
flows to a SFP  where either  $\Phi ( 0 ) = 2 \pi k$, or $\Phi ( 0 ) = \pi (2 k + 1)$
($k \in {\bf Z}$),
according to whether $\cos ( \theta )> 0 $, or $\cos ( \theta ) < 0$.
Nothing changes when $g>4$, since, even if
$G_2 ( L )$ becomes a relevant coupling, the
ratio $G_1 ( L ) / G_2 ( L )$ diverges,
as $L \to \infty$. At variance, when $B_\parallel = 0$,
the SFP is reached when  $\Phi ( 0 ) = 2 \pi k ,
{\bf S}^z = 1$, or $\Phi ( 0 ) = \pi (2 k + 1) ,
{\bf S}^z = - 1$ ($k$ integer).
In the following, we shall derive $\tilde{H}_{\bf B}$
which is the leading boundary perturbation,
at the SFP.

To construct  $\tilde{H}_{\bf B}$, one starts from the explicit
form of $\Phi ( x  ,\tau )$ obeying Dirichlet  BCs both at
$x=0$, and at $x=L$, since this BCs allow to determine the
zero-mode contribution. For this purpose, we need only to set
$\Phi ( L , \tau ) = \alpha$, since $\Phi ( 0 )$ is
determined by the boundary interaction. Thus, one
gets
\beq
\Phi ( x , \tau ) =   \alpha - \frac{1}{\sqrt{g}} \left[
\left( \frac{L-x}{L} \right)  \pi P \right] + \sqrt{\frac{2}{g}} \sum_{ n \neq 0}
\frac{a_n}{n} \: \sin \left[ \frac{\pi n x}{L} \right] \: e^{ - \frac{\pi n u \tau }{L} }
\:\:\:\: .
\label{sc1}
\eneq
\noindent
When $B_\parallel \neq 0$, the eigenvalues of
the zero-mode operator $P$ are either
given by $\sqrt{g} ( 2 k + \alpha / \pi)$, or by
$\sqrt{g} ( 2 k +1 + \alpha / \pi)$, according to
whether $\cos ( \theta )  > 0$, or $ \cos ( \theta )  < 0$;
at variance, when $\cos ( \theta )  = 0$, the
eigenvalues  are given by $p_k = \sqrt{g} ( k + \alpha / \pi)$.
Knowledge of the spectrum of $P$ allows for building the
leading boundary perturbation  at the SFP, using  the
delayed evaluation of boundary conditions  (DEBC) technique,
developed in Ref.\cite{aoc}.
Within the DEBC approach, a generic boundary perturbation at the SFP may
be represented as a linear combination of boundary vertex operators $V_{a,b} =
: \exp [ i ( a \Phi ( 0 ) + b \Theta ( 0 ) )]:$, with $\Theta ( x , \tau )$ being
the dual field of $\Phi ( x , \tau )$ \footnote{$\Theta ( x , \tau )$
is defined by the cross derivative relations,
$\frac{\partial \Phi ( x , \tau )}{ \partial x} = -\frac{i}{u} \frac{\partial \Theta ( x , \tau )}{
\partial \tau } $, and $\frac{\partial \Theta ( x , \tau )}{ \partial x} =
-\frac{i}{u} \frac{\partial \Phi ( x , \tau )}{
\partial \tau } $}, whose mode expansion is given by
\beq
\Theta ( x , \tau ) = \sqrt{g} \left[ \theta_0 + \frac{   \pi i u \tau}{L} P \right]
+ i \sqrt{2g}
\left\{ \sum_{ n \neq 0} \frac{a_n}{n} \: \cos \left[ \frac{ \pi n x}{L} \right] e^{ - \frac{ \pi n u \tau }{L}}
\right\}
\:\:\:\: .
\label{sc3.1}
\eneq
\noindent

Due to the  Dirichlet BCs
at $x=0$, $\Phi ( 0 )$ does not enter  $V_{a,b}$, and, from the commutator
 $[P , : \exp [ i b \Theta ( 0 ) ]:] =  2 b  \sqrt{g} : \exp [ i b \Theta ( 0 ) ]:$,
one sees that  $: \exp [ i b \Theta ( 0 ) ]:$ shifts $p_k$ to $p_k + 2 \sqrt{g}  b$.
Moreover, since  $p_k = \sqrt{g} ( k + \alpha / \pi)$, one gets $b =  m /2  $, $m \in {\bf Z}$.
As a result, in $\tilde{H}_{\bf B}$ there will be linear
combinations of  $ \cos \left[ m \frac{\Theta ( 0 )}{2} \right]$.
Since the symmetries of $H_{\bf B}$ require that
a shift in  $p_k$ is accompained by a change in the sign of
 ${\bf S}^z$, odd-$m$ terms must be multiplied by  ${\bf S}^x$.
As a result,  $\tilde{H}_{\bf B}$  may be written as

\beq
\tilde{H}_{\bf B} = -\mu_1 {\bf S}^x \cos \left[ \frac{\Theta ( 0 ) }{2} \right]
- \mu_2  \cos [ \Theta ( 0 ) ] - B_\perp {\bf S}^x - B_\parallel {\bf S}^z
\:\:\:\: .
\label{sc2}
\eneq
\noindent

Remarkably, the boundary interaction Hamiltonian $\tilde{H}_{\bf B}$ in
Eq.(\ref{sc2}) is the
same as $H_{\bf B}$ in Eq.(\ref{el2}),  provided that:

\beq
\Phi \leftrightarrow \Theta /2 \;\; , \;
{\bf S}^z  \leftrightarrow {\bf S}^x \;\; , \;
g_1 , g_2 \leftrightarrow
\mu_1 , \mu_2 \;\; , \;\;
B_\parallel  \leftrightarrow  B_\perp
\:\:\:\: .
\label{ssc2}
\eneq
\noindent
Thus, the RG analysis near by the SFP may carried out by mainly repeating
what has been already done near by the WFP.

Normal ordering of $\tilde{H}_{\bf B}$ is accounted for if one defines
 $\bar{\mu}_1 = \left( \frac{a}{L} \right)^{\frac{g}{4}}\: \mu_1$,
$\bar{\mu}_2 = \left( \frac{a}{L} \right)^{g} \: \mu_2$;
Furthermore, for $B_\perp=0$, repeating the same steps of   subsection
\ref{rg1}, yields the RG equations for the running coupling
strengths $\Lambda_1 ( L ) =
L \bar{\mu}_1$, and $\Lambda_2 ( L ) = L \bar{\mu}_2$, which are
given by

\begin{eqnarray}
 \frac{d \Lambda_1}{ d \ln \left( \frac{L}{L_0} \right)} &=&
\left[ 1 - \frac{g}{4} \right] \Lambda_1 + \Lambda_1 \Lambda_2 \nonumber \\
\frac{d \Lambda_2}{ d \ln \left( \frac{L}{L_0} \right)} &=&
\left[ 1 - g \right] \Lambda_2 + \frac{\Lambda_1 ^2}{2}
\:\:\:\: .
\label{sc3.a}
\end{eqnarray}
\noindent From Eqs.(\ref{ccompl2},\ref{ssc2}), one sees that, if
$B_\parallel \neq 0$, the vertex operators $: e^{ \pm \frac{i }{2}
\Theta ( 0 )}: $ are not a relevant perturbation to the SFP. These
operators may be regarded as the field theory representation of
the instanton/antiinstanton trajectories between minima of $H_{\rm
BDSG}$; in comparison with the instanton/antiinstanton
trajectories associated to the operators $: e^{ \pm i \Theta ( 0
)}: $, we refer to them as ``short instantons'' (SI). Since the
correlator $\langle {\bf T}_\tau [ : e^{ \frac{i  }{2} \Theta ( 0
)}: \: : e^{ - \frac{i  }{2} \Theta ( 0 )}: ] \rangle_{(0)}$ is always
accompained by a correlator of the local spin ${\bf S}^x$, one
sees that, for $B_\parallel \neq 0$ and $B_\perp = 0$, SIs are
always confined, and thus associated to a ``jump'' of $2 \pi$.  At
variance, for $1<g<4$ and $B_\parallel = 0$, SIs are deconfined,
allowing for tunneling between minima separated by $\pi$. Finally,
for $1<g<4$ and for $B_\parallel = B_\perp = 0$, the system flows
towards an attractive, IR stable finite FFP.

\begin{figure}
\includegraphics*[width=0.7\linewidth]{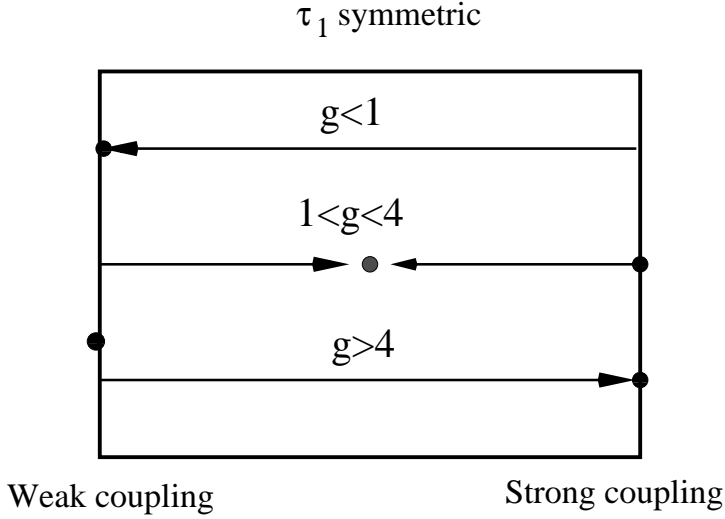}
\caption{ RG flow for $B_\parallel = 0 = B_\perp=0$. The boundary interaction
flows towards weak coupling for $g<1$, towards strong coupling for $g>4$, while,
for $1<g<4$, the RG flow points towards a finite coupling fixed point, represented
as a gray circle.}
\label{phased}
\end{figure}

\begin{figure}
\includegraphics*[width=1.1\linewidth]{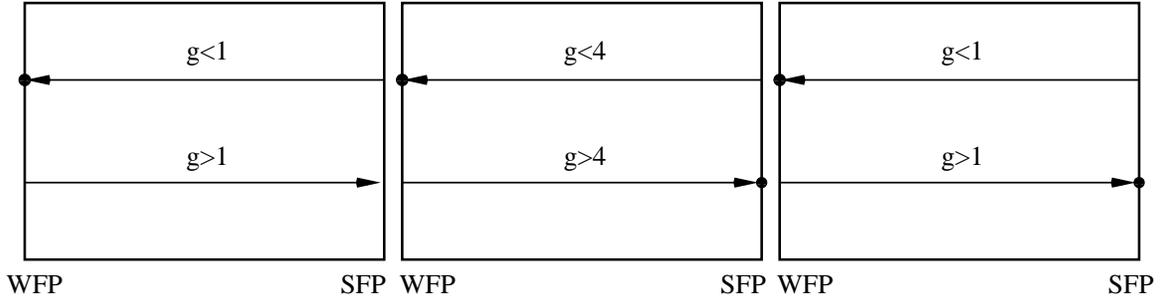}
\caption{RG flow diagram for various values of the Luttinger
parameter $g$ and of the control parameters $B_\parallel ,
B_\perp$: {\bf Left panel:} RG flow diagram for $B_\parallel
\neq 0 , B_\perp \neq 0$; {\bf Middle panel:} Flow diagram
for $B_\parallel = 0 , B_\perp \neq 0$; {\bf Right panel:}
Flow diagram for $B_\parallel \neq 0 , B_\perp = 0$. }
\label{phased2}
\end{figure}

We plot the phase diagram for $B_\perp = B_\parallel = 0$ in
Fig.\ref{phased}; there we show that the boundary interaction
flows towards weak coupling for $g < 1$, towards strong coupling
for $g > 4$ while, for $1 < g < 4$, it flows towards a finite
coupling fixed point. For the sake of completeness, the phase
diagrams corresponding to different values of $B_\parallel$ and
$B_\perp$ are reported in Fig.\ref{phased2}. From
Fig.\ref{phased2}, one sees that: for $B_\parallel \neq 0 ,
B_\perp \neq 0$, the boundary interaction is relevant or
irrelevant whether $g<1$, or $g>1$ and, since $B_\parallel \neq 0$
always confines the SIs near by the SFP, the WFP is IR stable for
$g<1$, while the SFP is IR stable only for $g>1$; at variance, for
$B_\parallel = 0 , B_\perp \neq 0$, charges $\pm 1$ are confined
over an imaginary time scale $\sim \lambda^{-1}$ (see
Eq.(\ref{ccompl2})), so that the leading effective boundary
perturbation at the WFP has scaling dimension $4/g$ and the SIs
are deconfined at the SFP. This makes the WFP IR stable for $g<4$,
and the SFP  IR stable for $g>4$; for $B_\parallel \neq 0 ,
B_\perp = 0$, one gets the same phase diagram as   for
$B_\parallel \neq 0, B_\perp \neq 0$.

SI deconfinement is the key mechanism destabilizing the SFP,
when the WFP is IR repulsive; for this reason, a detailed account
of this mechanism is provided in appendix \ref{instconf}. 
However, a similar - this time perturbative- mechanism holds near
the WFP since, as long as $B_\parallel \neq 0$, charges $\pm 1$
are always more likely to appear than charges $\pm 2$. As a
result, while for $g<1$ Coulomb gas charges are suppressed since
the corresponding fugacity is an irrelevant operator, for $1<g<4$,
charges $\pm 1$ proliferate, while the fugacity for charges $\pm
2$ still is irrelevant. Eventually, both charges may proliferate
for $g>4$, but the fugacity for charges $\pm 1$ is still more
relevant than the fugacity for charges $\pm 2$. Accordingly, for
$B_\parallel \neq 0$ and for $g>1$, the elementary charged
excitations tunnelling across {\bf C}  carry charge $\pm 1$ (in
units of the Cooper pair charge $e^*$). At variance, for
$B_\parallel = 0$, the fugacity $y_1$ goes to zero and only
charges $\pm 2$ go across {\bf C}, via coherent tunnelling of
pairs of Cooper pairs, induced by the boundary interaction $H_{\bf
B}$; it should be noticed that, in this range of parameters,
coherent tunnelling of Cooper pairs near the WFP emerges as a
perturbative phenomenon. We observe also that, for $B_\parallel =
0 ,  B_\perp  \neq 0$, the fugacity $y_1$ goes to zero and only charge
$\pm 2$ excitations cross {\bf C} as a result of the coherent
tunnelling  of pairs of Cooper pairs, induced by the boundary
interaction  $H_{\bf B}$.  Finally, for $B_\parallel = B_\perp =
0$, the fugacity $y_1$ is relevant for $g>1$ and the elementary
excitation carries charge $\pm 1$; as we shall discuss in more
detail in section \ref{dcjoscur},  for $B_\parallel = B_\perp =
0$, the JJN may be regarded as a "$4e$-deconfined superconductor",
since it exhibits $2e$ elementary charged excitations  while its
ground state correlations still support $4e$ superconductivity. 

In summary, our analysis evidences that a FFP emerges only when
the SIs are a relevant boundary perturbation to the SFP. As shown
in appendix \ref{instconf}, this happens only over length scales
$L$ such that $L_* \leq L \leq L_{\rm Stop} \sim L_0 \left(
\frac{1}{g_1 \cos ( \theta )}\right)$. For $B_\parallel \to 0$,
$L_{\rm Stop}  \to \infty$, and, thus, the FFP becomes a stable,
attractive fixed point.

\subsection{Frustration of decoherence near by the finite
coupling fixed point}
\label{rg3}

In this subsection, we focus on the analysis of the
interaction between the spin-1/2 degree of freedom sitting
at $x=0$ (${\bf S}$), and the plasmon modes of the leads,
which may be regarded as a bath of this two-level quantum
mechanical system. We shall show that, near by the FFP
(i.e., for $1<g<4$ and for $B_\parallel = 0$), the
decoherence induced by the plasmon bath is drastically
reduced  \cite{novais,giusonew}. To do this,
we compute the spectral density of states (SDOS) of
the local spin-1/2 variable, which is given by

\beq
S ( \Omega ) = \Im m \left[ \frac{ [ \chi^{\rm Tr} ] ( i \Omega + 0^+ ) }{ \Omega } \right]
\:\:\:\: ,
\label{fde1}
\eneq
\noindent
where $ [ \chi^{\rm Tr} ] ( \Omega )  $ is the transverse
spin susceptibility, defined as

\beq
 [ \chi^{\rm Tr} ] ( \Omega ) = \int_0^\infty \: d \tau \: e^{ i \Omega \tau}
\: \langle {\bf T}_\tau [ \{ \cos ( \theta ) {\bf S}_G^x ( \tau ) -
\sin ( \theta ) {\bf S}_G^z ( \tau ) \} \{ \cos ( \theta )
{\bf S}_G^x ( 0 ) - \sin ( \theta ) {\bf S}_G^z ( 0 ) \} ] \rangle
\:\:\:\: .
\label{fff1}
\eneq
\noindent
To carry out our task, it is most convenient to
compute  $[ \chi^{\rm Tr} ] ( \Omega )$ for $g=1,4$,
and then to extrapolate the behavior of $S ( \Omega )$
for any $1 <  g < 4$.

\begin{figure}
\includegraphics*[width=.85\linewidth]{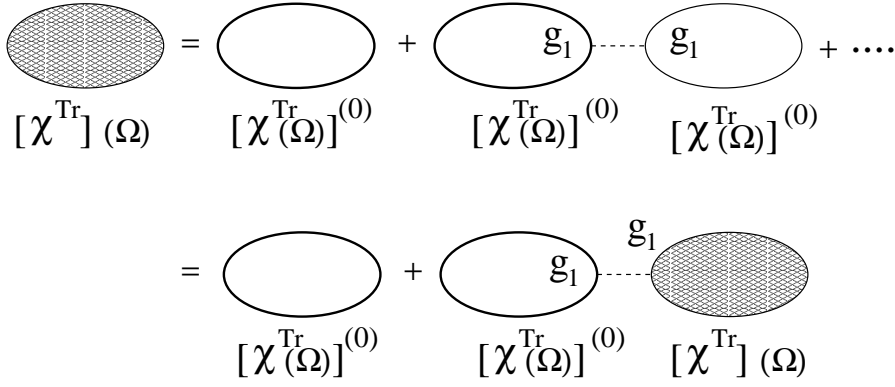}
\caption{Diagrams in the RPA sum yielding Eq.(\ref{ffp8}).}
\label{rpa}
\end{figure}

For $g=1$, $\cos [ 2 \Phi ( 0 ) ]$ is
an irrelevant perturbation and one may safely assume that
the FFP should lie at a distance $O ( g_1 )$ from the WFP,
so that Neumann BCs may be imposed on $\Phi $ at
$x= 0 $. As a result, one gets the simplified boundary
action given by

\beq
S_{\bf B} = S^{(0)}_{\bf B} - \frac{g_1}{2} \int_0^\beta \: d \tau \:
\{ a^\dagger a - b^\dagger b \} [ : e^{ i \Phi ( \tau  )} :   + : e^{ - i \Phi
( \tau ) }  : ]
\:\:\:\: ,
\label{ffp2}
\eneq
\noindent
with the operators $ : e^{ \pm i \Phi ( 0 )} : $
having scaling dimension equal to 1.
Similarly, for $g=4$, $ : e^{ \frac{i}{2} \Theta ( 0 )}: $ has
scaling dimension 1, and the Euclidean boundary action is
given by the dual of Eq.(\ref{ffp2}), namely

\beq
\tilde{S}_{\bf B} = S^{(0)}_{\bf B} - \frac{\mu_1}{2} \int_0^\beta \: d \tau \:
\{ a^\dagger b + b^\dagger a  \} [ : e^{ i \Theta ( \tau  ) } : 
+ : e^{ - i \Theta ( \tau ) } : ]
\:\:\:\: .
\label{ffp3}
\eneq
\noindent
By means of a Bogoliubov transformation to the normal modes
of the boundary fermions $\alpha , \beta$, one finds

\begin{eqnarray}
{\bf S}^z &=& \cos ( \theta ) \{ \alpha^\dagger \alpha - \beta^\dagger \beta \}
- \sin ( \theta ) \{ \alpha^\dagger \beta + \beta^\dagger \alpha \} \nonumber \\
{\bf S}^x &=& -\sin ( \theta ) \{ \alpha^\dagger \alpha - \beta^\dagger \beta \}
+ \cos ( \theta ) \{ \alpha^\dagger \beta + \beta^\dagger \alpha \}
\:\:\:\: ,
\label{ffp6}
\end{eqnarray}
\noindent
with
\beq
 \alpha = \cos ( \frac{\theta}{2} ) a + \sin ( \frac{\theta}{2} ) b \;\;\; , \;\;
\beta = - \sin ( \frac{\theta}{2} ) a + \cos ( \frac{\theta}{2} ) b
\:\:\:\: ,
\label{ffp4}
\eneq
\noindent
For $g=1$, one may resort to the random phase approximation (RPA)  used in
Ref.\cite{novais} to compute the ``dressed'' transverse spin susceptibility
(plotted in Fig.\ref{rpa}):
\beq
 \chi^{\rm Tr} ( \Omega )  = \frac{ [ \chi^{\rm Tr} ]^{(0)} ( \Omega ) }{1+
\frac{g_1^2 \sin^2 ( \theta )}{4} | \Omega |  [ \chi^{\rm Tr} ]^{(0)} ( \Omega ) }
\:\:\:\: ,
\label{ffp8}
\eneq
\noindent
where $[ \chi^{\rm Tr} ]^{(0)} ( \Omega )$ is given by

\[
 [ \chi^{\rm Tr} ]^{(0)} ( \Omega ) = \int_0^\infty \: d \tau \: e^{ i \Omega \tau}
\: \langle {\bf T}_\tau [ \{ \cos ( \theta ) {\bf S}^x ( \tau ) -
\sin ( \theta ) {\bf S}^z ( \tau ) \} \{ \cos ( \theta )
{\bf S}^x ( 0 ) - \sin ( \theta ) {\bf S}^z ( 0 ) \} ] \rangle_{(0)}
\]
\beq
= \frac{4 \lambda}{ \Omega^2 + 4 \lambda^2}
\:\:\:\: .
\label{fff1.b}
\eneq

From Eqs.(\ref{fde1},\ref{fff1.b}), one gets

\beq
S ( \Omega ) = \frac{ \Im m [ \chi^{\rm Tr} ( - i \Omega + 0^+ ) ]   }{\Omega }
= \frac{ 4 \lambda \frac{\sin^2 ( \theta ) g_1^2}{4}}{
( - \Omega^2 + 4 \lambda^2 )^2 + \left( \frac{g_1^2 \sin^2 ( \theta ) }{4} \right)
\Omega^2 }
\:\:\:\: .
\label{ffp9}
\eneq
\noindent

\begin{figure}
\includegraphics*[width=.7\linewidth]{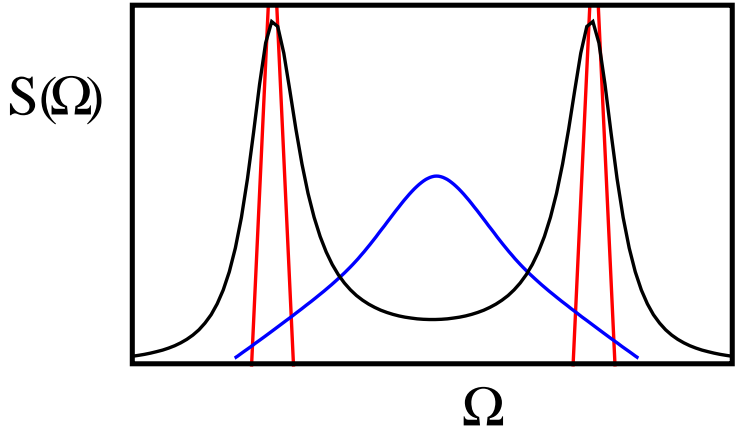}
\caption{Plot of the spectral density $S ( \Omega )$ in the various
regime accessible to the device: {\bf Blue line:} $ S ( \Omega )$ near by
the WFP; {\bf Red line:} $ S ( \Omega )$ near by
the SFP; {\bf Black line:} $ S ( \Omega )$ near by
the FFP (see Ref.\cite{giusonew} for details).}
\label{plot}
\end{figure}

A plot of $ S ( \Omega ) $ {\it vs.} $\Omega$ is reported in
Fig.\ref{plot}. One sees that the limited broadening of the two
peaks at $\Omega \sim 2 \lambda$ (which is the signal of
frustration of decoherence) depends on having both $g_1$ finite,
and $\sin ( \theta ) \neq 0$. Furthermore, choosing $g=1$, implies
that $g_1$ is the coefficient of a marginal perturbation and,
thus, is not renormalized by  the interaction.

For  $g>1$, the system is attracted by an IR stable FFP and
one has that $g_1$ goes to $g_1^*$, where  $g_1^*$ is the value of
this coupling at the FFP.  Even if in section \ref{rengrou} it has
been shown that the FFP is IR attractive only when $B_\parallel =
B_\perp = 0$, the evaluation of  the"dressed"transverse spin
susceptibility requires to slightly move from the FFP by applying
a small, nonzero $B_\parallel$ and/or $B_\perp$, in order to split
the two impurity levels \cite{novais}. This may be safely
carried out when the JJN has a finite size L since, in this case,
the FFP is stable also against small fluctuations of the control
parameters $B_\parallel , B_\perp$, provided that $u/L$ is
sufficiently big (see appendix \ref{instconf}). As a result, for
$g>1$,  the equation yielding $S ( \Omega )$ has the same form as
Eq.(\ref{ffp9}), provided one replaces $g_1$ with  $g_1^*$ and
introduces  a ``self-energy correction'' $\propto  | \Omega |^{
\frac{2}{g} - 1}$ 
 \cite{giusonew}. From the duality relations given in Eq.(\ref{ssc2}), one finds
that Eq.(\ref{ffp9}) holds also for $g=4$. Our results confirm
that frustration of decoherence may be a remarkable signature of
the emergence of a FFP in the phase diagram accessible to
superconducting quantum devices \cite{giusonew}.

\section{Josephson current and coherent tunneling of Cooper pairs}
\label{dcjoscur}

In section \ref{model}, we showed that the Hamiltonian $H_{\bf B}$
is invariant under the $\tau_1$-symmetry, even if it contains a
term $\propto \cos [ \Phi ( 0 ) ]$. In the following, we show that
this symmetry is responsible for the appearance of $4e$
superconducting correlations  in all the phases accessible to the
JJN. For this purpose, we shall compute the Josephson current (JC)
across {\bf C} in all the phases of the JJN.

To induce a JC, one may connect the leads
of the JJN to two bulk superconductors
at fixed phase difference $\alpha$ (see Fig.\ref{device})
\cite{glark,giuso2}. This amounts to require that, at
the outer boundary ($x=L$),  $\Phi ( L , t ) = \alpha$.
The (0-temperature)  dc JC  $ I ( \alpha )$ may then be computed as

\beq
I ( \alpha ) = - \lim_{\beta \to \infty}  \frac{e^*}{\beta}
\frac{\partial \ln {\bf Z} }{ \partial \alpha}
\:\:\:\: ,
\label{dcjos1.a}
\eneq
\noindent
with $e^*=2 e$ being the Cooper pair charge and
${\bf Z} $ defined in Eq.(\ref{compl0.2})

\subsection{DC Josephson current near by the weakly coupled fixed point}
\label{wdcj}

\begin{figure}
\includegraphics*[width=.7\linewidth]{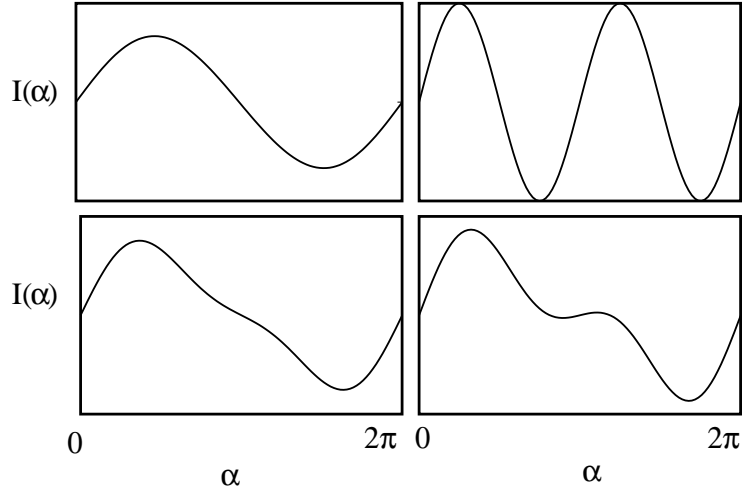}
\caption{Plot of the dc Josephson current across {\bf C} at the
WFP for $\bar{g}_2 / \bar{g}_1 = .3$ and for different
values of $B_\parallel / B_\perp$. {\bf a):} Top left panel:
$B_\parallel / B_\perp = 10$; {\bf b):} Bottom left panel:
$B_\parallel / B_\perp = .1$;
 {\bf c):} Bottom right panel: $B_\parallel / B_\perp = .05$;
{\bf d):} Top right panel: $B_\parallel / B_\perp = 0$.}
 \label{joswc}
\end{figure}

Near by the WFP, $\Phi ( x , \tau )$ is given by

\beq
\Phi ( x , \tau ) = \alpha +
\sum_{ n \in {\bf Z}} \frac{\alpha ( n ) }{ n + \frac{1}{2}} \cos \left[
\left( n + \frac{1}{2} \right) \frac{\pi x}{L} \right] e^{ -
\left( n + \frac{1}{2} \right)\frac{ \pi  u \tau}{L}  }
\:\:\:\: ,
\label{dcjos1}
\eneq
\noindent
with
\[
  [ \alpha ( n ) , \alpha (n' ) ]
=  \delta_{n + n' - 1 } \left( n + \frac{1}{2} \right)
\:\:\:\: .
\]
\noindent
Taking into account that
$\langle :e^{ i a \Phi ( 0 ,  \tau )}: \rangle_{(0)}  = e^{ i a \alpha}$,
one gets

\[
\ln {\bf Z} \approx \ln {\bf Z}_0 +  y_1   \int_0^\beta \; d \tau \:
\langle : \cos \left[ \Phi  ( 0 , \tau )
\right] : \rangle_{(0)} +  y_2 \int_0^\beta \; d \tau \:
\langle : \cos \left[ 2  \Phi ( 0 , \tau )
\right] :  \rangle_{(0)}
\]
\beq
= \ln {\bf Z}_0 +  \beta \{ y_1   \cos (\alpha )
+  y_2  \cos (2 \alpha )\}
\:\:\:\: ,
\label{dcjos2}
\eneq
\noindent
from which the dc JC is given by

\beq
I ( \alpha ) \approx e^* y_1  \sin ( \alpha ) +
2 e^* y_2 \sin ( 2 \alpha )
\;\;\;\; .
\label{dcjos3}
\eneq
\noindent
By inspection of Eq.(\ref{dcjos3}), one sees   that $I ( \alpha )$
is the sum of a term ($\propto \sin (\alpha)$) corresponding
to the usual tunneling of Cooper pairs (of charge $e^*$) across {\bf C}, and of
a term ($\propto \sin ( 2 \alpha)$) describing
the coherent tunneling of pairs of Cooper pairs (CTCP) (of
total charge $2e^*$) across {\bf C}. The relative weight of
the two terms contributing to $I ( \alpha )$
is $\propto \frac{y_1}{y_2} \propto \cos ( \theta )$.
 Thus, it may be tuned upon acting on   $\varphi$  till, at
$\varphi = \pi$, only the term $\propto \sin ( 2 \alpha )$ is left, that is,
the JC flows across {\bf C} only because of CTCP. In Fig.\ref{joswc}, we
plot the dc JC across {\bf C} for different values of $\varphi$.

The CTCP is also evidenced in the ac JC, arising across {\bf C}
when a finite voltage bias $V$ is applied to the ends of
both leads. To account for $V$, one has simply to add
to $H_{\rm BDSG}$ a  ``voltage bias'' term
$H_V = \frac{g e^*V}{2 \pi} \frac{\partial \Phi}{\partial t}$,
which yields a time-dependent shift of
$\Phi$ as $\Phi \to \Phi + e^* V t$, from which
the ac JC is given by

\beq
I ( \alpha , t  ) \to I ( \alpha + e^* V t ) = e^* y_1  \sin ( \alpha + e^* V t ) +
2 e^* y_2 \sin ( 2 \alpha + 2 e^* V t )
\;\;\;\; .
\label{dcjos4}
\eneq
\noindent

\subsection{DC Josephson current near by the strongly coupled fixed point}
\label{sdcj}

Near by the SFP, the field $\Phi$, given by Eq.(\ref{scj1}), contains
contributions from the zero mode operator. As a result, the partition
function may be factorized as  ${\bf Z} = {\bf Z}_{\rm 0-mode} [ \alpha ]
{\bf Z}_{\rm osc}$, where ${\bf Z}_{\rm 0-mode} [ \alpha ]$ is
the contribution of the zero-mode operators.
From the analysis of the zero mode spectrum carried in subsection
\ref{rg2}, one sees that, for $B_\parallel \geq 0$,
the eigenvalues of the  zero-mode operator
$P$ are given by $\{ p_k \} = \{ \sqrt{g} (k +
\alpha / \pi ) \}$, with the odd-$k$ eigenvalues higher in
energy by $\Delta \epsilon = \bar{g}_1 \cos ( \theta ) $. As
a result
\begin{figure}
\includegraphics*[width=.7\linewidth]{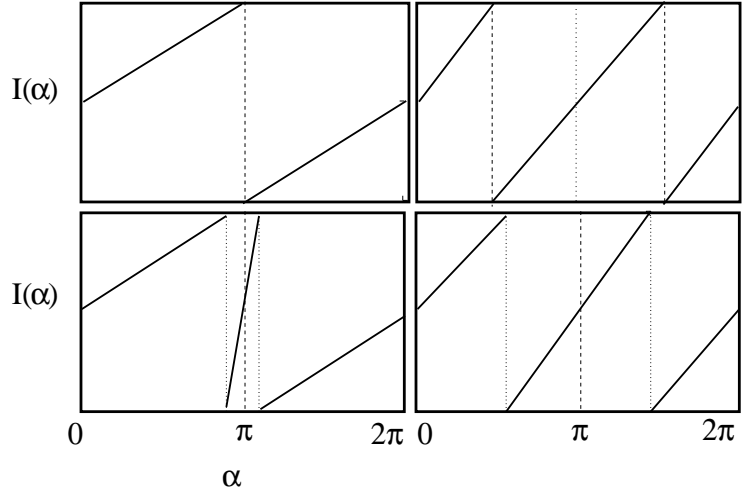}
\caption{Plot of the dc Josephson current across {\bf C} at the
SFP for (from top left panel to top right panel, rotating
counterclockwisely) $\Delta \epsilon/\frac{\pi u }{4L} \gg 1$;
$\Delta \epsilon/\frac{\pi u }{4L} \sim 1$; $\Delta
\epsilon/\frac{\pi u }{4L} \ll 1$; $\Delta \epsilon/\frac{\pi u
}{4L} = 0$.}
 \label{jossc}
\end{figure}
\beq
{\bf Z}_{\rm 0-mode} [ \alpha ]
= \sum_{ k \in Z} \: \exp \left[ - \beta \left( \frac{\pi u }{4L} ( p_k )^2 + [1 - (-1)^k]
\frac{\Delta \epsilon}{2} \right) \right]
\:\:\:\: ,
\label{scj1}
\eneq
\noindent
from which the dc JC is obtained as
\beq
I ( \alpha ) = \lim_{\beta \to \infty} \frac{e^*}{\beta}
\frac{\partial \ln {\bf Z}_{\rm 0-mode} [ \alpha ] }{ \partial \alpha}
\:\:\:\: .
\label{scj2}
\eneq
\noindent
In Fig.\ref{jossc} we plot $I ( \alpha )$ {\it vs.} $\alpha$
for different values of  $\Delta \epsilon/\frac{\pi u }{4L} $.
One notices that  for $\Delta \epsilon/\frac{\pi u }{4L}
\gg 1 $, the current takes the usual sawtooth behavior with periodicity in $\alpha$
equal to $2 \pi$. Furthermore, as $\Delta \epsilon/\frac{\pi u }{4L} \sim 1$,
one sees that the current continuously crosses
$\alpha = \pi$, with ``satellite'' jumps at $\alpha = \pi \pm \left[
\frac{\pi}{2} - \frac{4L \Delta \epsilon}{\pi u g} \right]$. 
As $\Delta \epsilon/\frac{\pi u }{4L}$
approaches 0, the satellite jumps move to $\pi \pm \pi / 2$, where
they eventually take place, when $\Delta \epsilon = 0$.
As a result, the period of the sawtooth
is exactly halved at $\Delta \epsilon/\frac{\pi u }{4L} = 0$.

\subsection{DC Josephson current near by the finite coupling
fixed point}
\label{fdjc}

To compute the JC near by the FFP, one must include the effects of
the relevant SI perturbation at the SFP. While long instantons
just provide a smoothing of the spikes of $I (\alpha)$ at the SFP
\cite{giusonew}, for $B_\parallel = 0$ {\bf and $B_\perp =0$}, SIs
become a relevant perturbation and drive the system to the IR
attractive FFP. As a result, for $B_\parallel = 0$ {\bf and
$B_\perp =0$}, one expects that the net effect of the SIs on the
dc JC will be to smoothen the sharp jumps of the sawtooth depicted
in Fig.\ref{jossc}, while still preserving the half periodicity of
$ I ( \alpha )$.

\begin{figure}
\includegraphics*[width=.7\linewidth]{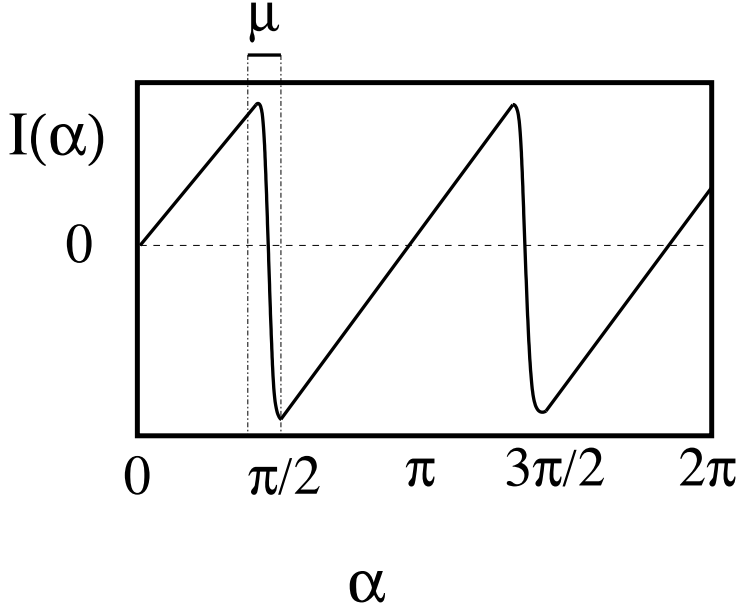}
\caption{Sketch of $I ( \alpha )$ {\it vs.} $\alpha$ near
by the FFP (Eq.(\ref{fjc5})).}
 \label{josfc}
\end{figure}

In the following, we compute $ I ( \alpha )$ near by the
jump at $\alpha = \pi /2$ for $\Delta \epsilon/\frac{\pi u }{4L} = 0$
(top right panel of Fig.\ref{jossc}). Upon tuning $\alpha$ so
that $0 < \alpha < \pi$, the low energy
states contributing to $ I ( \alpha )$  are
 associated with the zero modes labeled by
$p_k = p_0$ and $p_k = - 1$. Their energies
are respectively given by $E_0 ( \alpha )
\pm \epsilon_0 ( \alpha )$, with

\beq
E_0 ( \alpha ) = \frac{g}{4 \pi L} \left[ \frac{\pi^2}{4} +
\left( \frac{\pi}{2} - \alpha \right)^2 \right] \;\;\; , \;\;
\epsilon_0 ( \alpha ) = \frac{g}{4 \pi L} \left( \frac{\pi }{2} -
\alpha \right)
\:\:\:\: .
\label{fjc1}
\eneq
\noindent
Since, for $p=p_0$, ${\bf S}^z = 1$, while,
$p=p_{-1}$, ${\bf S}^z = -1$, one
may trade the degeneracy-breaking energy $\epsilon_0 ( \alpha)$
with  an effective $B_\parallel ( \alpha ) = \epsilon_0  (\alpha )$;
in addition, as one is approaching the FFP from the SFP, one
should take,  as boundary interaction, the dual boundary
Hamiltonian $\tilde{H}_{\bf B}$ given in Eq.(\ref{sc2});
the term ($\propto \cos [ \Theta ( 0 ) / 2]$) now acts on
the spin ${\bf S}$ as an effective field
$B_\perp \sim \bar{\mu}_1$.  As a result,
one may compute
$ I ( \alpha )$ for $\alpha \sim \pi / 2$ by taking
the logarithmic derivative with respect to $\alpha$ of
the relevant contribution to ${\bf Z}_{\rm 0-mode} [ \alpha ]$;
namely

\beq
{\bf Z}_{\rm 0-mode} [ \alpha ]  \sim e^{ - \beta E_0 ( \alpha )} \int \:{\bf D} \{ a , b \} \:
e^{ - S_E [ \{ a , b \} ] + \ldots }
\:\:\:\: ,
\label{fjc2}
\eneq
\noindent
with

\beq
 S_E [ \{ a , b \} ] = \int_0^\beta \: d \tau \: \left\{ a^\dagger \left[
\frac{\partial }{ \partial \tau} - B_\parallel ( \alpha ) \right] a +
b^\dagger \left[
\frac{\partial }{ \partial \tau} + B_\parallel ( \alpha ) \right] b
+ B_\perp [ a^\dagger b + b^\dagger a ] \right\}
\:\:\:\: ,
\label{fjc3}
\eneq
\noindent
while the ellipses denote contributions coming from $p_k \neq \{ p_0 , p_{-1} \}$.
Using Eq.(\ref{dcjos1.a}), one gets

\beq
I ( \alpha ) \approx - \frac{e^* g}{ 2 \pi L} \left( \alpha - \frac{\pi}{2} \right)
\left\{ 1 + \frac{ \left( \alpha - \frac{\pi}{2} \right)}{ \left( \alpha -
\frac{\pi}{2} \right)^2 + \left( \frac{ 4 \pi \mu_1}{g} \right)^2 L^{ 2 - \frac{g}{2}} }\right\}
\:\:\:\: .
\label{fjc5}
\eneq
\noindent
Eq.(\ref{fjc5}) shows that $ I ( \alpha )$ is a smooth function,
linearly crossing 0, at $\alpha = \pi / 2$. The term
 $\propto L^{2-\frac{g}{2}}$ in the denominator  evidences that
this behavior is more pronounced, as the system size
increases. Carrying out a similar computation for
$\alpha \neq \pi / 2$ leads to the plot displayed in
Fig.\ref{josfc}.

In the next section, we shall show how the results obtained
for the coherent current transport in the JJN extend to the
situation in which a dissipative current is induced, as well.

\section{DC transport and coherent tunneling of Cooper pairs}
\label{dctra1}

In this section, we analyze the observable consequences of
CTCP on a dc transport experiment. For this purpose, we shall
compute the dissipative current $I_{\rm DC} (V)$ flowing across
{\bf C} when the leads are connected to two metallic
wires, at fixed voltage bias $V$. Differently from the Josephson supercurrent,
$I_{\rm DC} (V)$ fluctuates even at $T= 0$, in presence of
weak tunneling and/or backscattering centers, and this
zero-temperature fluctuations are measured by
the shot noise at voltage bias $V$ and frequency $\Omega$,

\beq
S ( V , \Omega ) = \int \: d t \: e^{ i \Omega t}
 \langle \{ I_{\rm fl}  ( t ) , I_{\rm fl} ( 0 ) \} \rangle_{\rm NE}
\:\:\:\: ,
\label{sshot1}
\eneq
\noindent
 where $\{ , \}$ is the anticommutator,  $\langle \ldots \rangle_{\rm NE}$
denotes averaging, at zero temperature and at finite  $V$,
 $I_{\rm fl}$ is either identified
with  the transmitted current,  when the two leads are weakly
coupled to each other, or with the backscattered
current, when the leads are strongly coupled.

Measuring the shot noise provides a tool to probe  the effective
elementary charge $q$ flowing across a device since, from the
functional dependence of $S ( V , \Omega )$ upon $V$, one obtains
$q$ as

\beq
 q ( V ) = S ( V , \Omega \to 0) / ( 2 I_{\rm fl} ( V ))
\:\:\:\: .
\label{sshot2}
\eneq
\noindent
Measurements of $S$ have been
used to probe noninteger charges in fractional
quantum Hall bars \cite{noisehallexpe}, in normal metal-superconductor
junctions \cite{nms}, in Kondo dots \cite{sela}. In the following,
we show that, by properly tuning the external control parameter,
pairing of Cooper pairs may be  as well detected in a shot noise measurement,
provided that $V$ is lower than a threshold, depending
on the values of $B_\parallel $ and $B_\perp$.

In subsection \ref{sno1},
we compute $I_{\rm fl} ( V ) $ and $S ( V )$ at finite $V$ near by the WFP.
We show that, when $B_\parallel = 0$,
 $B_\perp$ defines a threshold below which
the effective charge detected from a shot noise measurement is
$q = 2e^*$, which evidences a phase with
$4e$ superconductivity.

In subsection \ref{sno2}, we use the fermionization
procedure of the BDSG Hamiltonian for $g=2$ \cite{ameduri},
to extend the results of subsection \ref{sno1} beyond the perturbative
regime in the boundary couplings. Mathematical
details concerning the derivation of subsection \ref{sno1} are
 provided in appendix \ref{integr}, while the fermionization
procedure is discussed in detail in appendix \ref{shot}.

\subsection{Perturbative computation of the dissipative current and of the
shot noise at finite voltage bias $V$}
\label{sno1}

When  the JJN is connected to two  metallic wires at
finite voltage bias $V$, the computation of the
current flowing across {\bf C} -and its fluctuations-  may
be carried out within the  Hamiltonian formalism.
In particular, near by the WFP, one may
consistently compute both $I_{\rm DC} ( V ) $ and $ S ( V )$
nonperturbatively in $V$, resorting to time-dependent
perturbation theory in the boundary couplings $g_1$, $g_2$.

Applying a finite voltage bias $V$ amounts to shift
the field $\Phi$ as $\Phi \to \Phi + \omega_V t$, with
$\omega_V = e^* V$.  To the leading order in $g_1 , g_2$,
the current operator in the Heisenberg
representation,  $j_H  ( t ) $, is obtained  by
\beq
 j_H ( t ) \approx j_I ( t ) + i \int_{-\infty}^t \: d \tau \: [ H_{\bf B} ( \tau ) ,
j_I ( t ) ]
\:\:\:\: ,
\label{cicciopa8}
\eneq
\noindent
where $j_I ( t ) $ and $ H_{\bf B} ( t ) $ are, respectively, the
current operator and the boundary Hamiltonian in the interaction
representation, given by
\begin{eqnarray}
 j_I ( t ) &=& e^* \bar{g}_1 {\bf S}^z ( t ) : \sin [ \Phi ( t ) + \omega_V t ] : +
2 e^* \bar{g}_2  : \sin [ 2 \Phi ( t ) + 2 \omega_V t ]  : \nonumber \\
H_{\bf B} ( t ) &=& - \bar{g}_1 {\bf S}^z ( t ) : \cos [ \Phi ( t ) + \omega_V t ]: -
 \bar{g}_2  : \cos [ 2 \Phi ( t ) + 2 \omega_V t ] :
\:\:\:\: ,
\label{cicciopabis1}
\end{eqnarray}
\noindent
with $\Phi ( t ) \equiv \Phi ( x = 0 , t )$.

The 0-temperature  average value
of  $ j_H ( t ) $ is given by
\[
 I_{\rm DC} ( V ) = - e^* \cos^2 ( \theta ) \bar{g}_1^2 \: \Re e \left\{ \int_0^\infty \:
d \tau \: \frac{ [ e^{ i \omega_V \tau} - e^{ - i \omega_V \tau} ]}{ [
e^{ \frac{\pi i u}{L} \tau} - e^{ - \frac{\pi i u}{L} (\tau - i \eta ) } ]^{\frac{2}{g}}}
\right\}
\]
\[
-  e^* \sin^2 ( \theta ) \bar{g}_1^2 \: \Re e \left\{ \int_0^\infty \:
d \tau \: \frac{ e^{ - 2 i \lambda \tau}  [ e^{ i \omega_V \tau} - e^{ - i \omega_V \tau} ]}{ [
e^{ \frac{\pi i u}{L} \tau} - e^{ - \frac{\pi i u}{L} (\tau - i \eta ) } ]^{\frac{2}{g}}}
\right\}
\]
\beq
- 2 e^*  \bar{g}_2^2 \: \Re e \left\{ \int_0^\infty \:
d \tau \: \frac{   [ e^{ 2 i \omega_V \tau} - e^{ - 2 i \omega_V \tau} ]}{ [
e^{ \frac{\pi i u}{L} \tau} - e^{ - \frac{\pi i u}{L} (\tau - i \eta ) } ]^{\frac{8}{g}}}
\right\}
\:\:\:\: ,
\label{sper1}
\eneq
\noindent
while $S ( V )$ is given by

\[
S ( V ) = - 2 (e^* )^2 \cos^2 ( \theta ) \bar{g}_1^2 \: \Re e \left\{ \int_0^\infty \:
d \tau \: \frac{ [ e^{ i \omega_V \tau} + e^{ - i \omega_V \tau} ]}{ [
e^{ \frac{\pi i u}{L} \tau} - e^{ - \frac{\pi i u}{L} (\tau - i \eta ) } ]^{\frac{2}{g}}}
\right\}
\]
\[
- 2  (e^*)^2 \sin^2 ( \theta ) \bar{g}_1^2 \: \Re e \left\{ \int_0^\infty \:
d \tau \: \frac{ e^{ - 2 i \lambda \tau}  [ e^{ i \omega_V \tau} + e^{ - i \omega_V \tau} ]}{ [
e^{ \frac{\pi i u}{L} \tau} - e^{ - \frac{\pi i u}{L} (\tau - i \eta ) } ]^{\frac{2}{g}}}
\right\}
\]
\beq
- 8 (e^*)^2  \bar{g}_2^2 \: \Re e \left\{ \int_0^\infty \:
d \tau \: \frac{   [ e^{ 2 i \omega_V \tau} + e^{ - 2 i \omega_V \tau} ]}{ [
e^{ \frac{\pi i u}{L} \tau} - e^{ - \frac{\pi i u}{L} (\tau - i \eta ) } ]^{\frac{8}{g}}}
\right\}
\:\:\:\: .
\label{sper2}
\eneq
\noindent
The relevant integrals needed in Eqs.(\ref{sper1},\ref{sper2}) are
computed in  appendix \ref{integr}. As a result,  $I_{\rm DC} ( V )$ and
$ S ( V )$ are given by

\[
 I_{\rm DC} ( V ) =  e^* \cos^2 ( \theta ) \bar{g}_1^2 \: \Re e \biggl\{i
\biggl[ \frac{ \Gamma [ 1 - \frac{2}{g} ] \: \Gamma [ \frac{L \omega_V}{2 \pi u} +
\frac{1}{g} + 1 ]}{ \left( \omega_V - \frac{2 \pi u}{g L} \right)
\Gamma [  \frac{L \omega_V}{2 \pi u}
-\frac{1}{g} + 1 ]} +  \frac{ \Gamma [ 1 - \frac{2}{g} ] \: \Gamma [ - \frac{L \omega_V}{2 \pi u} +
\frac{1}{g} + 1 ]}{ \left( \omega_V + \frac{2 \pi u}{g L} \right)
\Gamma [ - \frac{L \omega_V}{2 \pi u} 
-\frac{1}{g} + 1 ]} \biggr] \biggr\}
\]
\[
 +  e^* \bar{g}_1^2 \sin^2 ( \theta ) \: \Re e \biggl\{i
\biggl[ \frac{ \Gamma [ 1 - \frac{2}{g} ] \: \Gamma [ \frac{L (\omega_V + 2 \lambda ) }{2 \pi u} +
\frac{1}{g} + 1 ]}{ \left(  \omega_V + 2 \lambda - \frac{2 \pi u}{g L} \right)
\Gamma [  \frac{L (\omega_V + 2 \lambda)}{2 \pi u}
-\frac{1}{g} + 1 ]}
\]
\[
 +  \frac{ \Gamma [ 1 - \frac{2}{g} ] \: \Gamma [ - \frac{L (\omega_V
- 2 \lambda )}{2 \pi u}
 + \frac{1}{g} + 1 ]}{ \left( \omega_V -2 \lambda+ \frac{2 \pi u}{g L} \right)
\Gamma [ - \frac{L (\omega_V - 2 \lambda)}{2 \pi u} 
-\frac{1}{g} + 1 ]}\biggr] \biggr\}
\]
\beq
+ 2 e^* \bar{g}_2^2 \: \Re e \biggl\{i
\biggl[ \frac{ \Gamma [ 1 - \frac{8}{g} ] \: \Gamma [ \frac{L \omega_V}{ \pi u} +
\frac{4}{g} + 1 ]}{ \left( 2 \omega_V - \frac{8 \pi u}{g L} \right)
\Gamma [  \frac{L \omega_V}{ \pi u}
-\frac{4}{g} + 1 ]} +  \frac{ \Gamma [ 1 - \frac{8}{g} ] \: \Gamma [ - \frac{L \omega_V}{ \pi u} +
\frac{4}{g} + 1 ]}{ \left( 2 \omega_V + \frac{8 \pi u}{g L} \right)
\Gamma [ - \frac{L \omega_V}{ \pi u} 
-\frac{4}{g} + 1 ]}\biggr] \biggr\}
\:\:\:\: ,
\label{sper3}
\eneq
\noindent
and

\[
S( V ) =  2 (e^*)^2 \cos^2 ( \theta ) \bar{g}_1^2 \: \Re e \biggl\{i
\biggl[ - \frac{ \Gamma [ 1 - \frac{2}{g} ] \: \Gamma [ \frac{L \omega_V}{2 \pi u} +
\frac{1}{g} + 1 ]}{ \left( \omega_V - \frac{2 \pi u}{g L} \right)
\Gamma [  \frac{L \omega_V}{2 \pi u}
- \frac{1}{g} + 1 ]} -  \frac{ \Gamma [ 1 - \frac{2}{g} ] \: \Gamma [ - \frac{L \omega_V}{2 \pi u} +
\frac{1}{g} + 1 ]}{ \left( \omega_V + \frac{2 \pi u}{g L} \right)
\Gamma [ - \frac{L \omega_V}{2 \pi u} 
-\frac{1}{g} + 1 ]}\biggr] \biggr\}
\]
\[
 + 2 ( e^* )^2 \bar{g}_1^2 \sin^2 ( \theta ) \: \Re e \biggl\{i
\biggl[ - \frac{ \Gamma [ 1 - \frac{2}{g} ] \: \Gamma [ \frac{L (\omega_V + 2 \lambda ) }{2 \pi u} +
\frac{1}{g} + 1 ]}{ \left(  \omega_V + 2 \lambda - \frac{2 \pi u}{g L} \right)
\Gamma [  \frac{L (\omega_V +  2 \lambda)}{2 \pi u} 
-\frac{1}{g} + 1 ]}
\]
\[
 +  \frac{ \Gamma [ 1 - \frac{2}{g} ] \: \Gamma [ - \frac{L (\omega_V
- 2 \lambda )}{2 \pi u} +
\frac{1}{g} + 1 ]}{ \left( \omega_V  - 2 \lambda+ \frac{2 \pi u}{g L} \right)
\Gamma [ - \frac{L (\omega_V-  2 \lambda)}{2 \pi u} 
-\frac{1}{g} + 1 ]}\biggr] \biggr\}
\]
\beq
+ 8 ( e^* )^2 \bar{g}_2^2 \: \Re e \biggl\{i
\biggl[ - \frac{ \Gamma [ 1 - \frac{8}{g} ] \: \Gamma [ \frac{L \omega_V}{ \pi u} +
\frac{4}{g} + 1 ]}{ \left( 2 \omega_V - \frac{8 \pi u}{g L} \right)
\Gamma [  \frac{L \omega_V}{ \pi u}
-\frac{4}{g} + 1 ]} +  \frac{ \Gamma [ 1 - \frac{8}{g} ] \: \Gamma [ - \frac{L \omega_V}{ \pi u} +
\frac{4}{g} + 1 ]}{ \left( 2 \omega_V + \frac{8 \pi u}{g L} \right)
\Gamma [ - \frac{L \omega_V}{ \pi u} 
-\frac{4}{g} + 1 ]}\biggr] \biggr\}
\:\:\:\: .
\label{sper4}
\eneq
\noindent
Finally, the Stirling's approximation (Eq.(\ref{iden6})) allows to derive
the large-$L$ limit ($( L \omega_V / 2 \pi u )
\gg 1$) of Eqs.(\ref{sper3},\ref{sper4}), leading to

\[
 I_{\rm DC} ( V ) \approx e^* \Gamma [ 1 - \frac{2}{g} ] \sin \left( \frac{2 \pi}{g} \right)
\biggl\{ \frac{ \bar{g}_1^2 \cos^2 ( \theta ) }{\omega_V} \left( \frac{L \omega_V}{2 \pi u } \right)^{\frac{2}{g} }
\]
\beq
+ \frac{ \bar{g}_1^2\sin^2 ( \theta ) }{ \omega_V - 2 \lambda}
\: \theta ( \omega_V - 2 \lambda ) \left( \frac{L ( \omega_V - 2 \lambda ) }{2 \pi u }
\right)^\frac{2}{g} \biggr\} + 2 e^* \Gamma [ 1 - \frac{8}{g} ] \sin \left( \frac{8 \pi}{g} \right)
\frac{ \bar{g}_2^2}{ 2 \omega_V} \left( \frac{L \omega_V}{ \pi u } \right)^\frac{8}{g}
\:\:\:\: ,
\label{sper5}
\eneq
\noindent
 and

\[
S( V ) \approx 2 ( e^* )^2 \Gamma [ 1 - \frac{2}{g} ] \sin \left( \frac{2 \pi}{g} \right)
\biggl\{ \frac{ \bar{g}_1^2 \cos^2 ( \theta ) }{\omega_V} \left( \frac{L \omega_V}{2 \pi u } \right)^{\frac{2}{g} }
\]
\beq
+ \frac{ \bar{g}_1^2\sin^2 ( \theta ) }{ \omega_V - 2 \lambda}
\: \theta ( \omega_V - 2 \lambda ) \left( \frac{L ( \omega_V - 2 \lambda ) }{2 \pi u }
\right)^\frac{2}{g} \biggr\} + 8 ( e^* )^2 \Gamma [ 1 - \frac{8}{g} ] \sin \left( \frac{8 \pi}{g} \right)
\frac{ \bar{g}_2^2}{ 2 \omega_V} \left( \frac{L \omega_V}{ \pi u } \right)^\frac{8}{g}
\:\:\:\: ,
\label{sper6}
\eneq
\noindent
with $\theta ( x ) $ being Heavside's $\theta$-function.

To determine $q ( V)$, one observes that
the term $\propto \bar{g}_1^2 \sin^2 ( \theta )$ in
Eqs.(\ref{sper5},\ref{sper6}) exhibits a threshold at $\omega_V = 2 \lambda$.
In computing $q ( V )$, one notices that, since the term $\propto \cos [ \Phi ( 0 ) ]$
and the term $\propto \cos [ 2 \Phi ( 0 ) ]$ in
 $H_{\bf B}$ have different scaling  dimensions  ($1/g$ and $4/g$, respectively),
tuning the low-energy  scale $\omega_V$ induces a flow in
$q ( V ) $. Indeed, assuming $B_\parallel  \neq 0$, one finds
that $q (V)  \sim e^*$ for $\omega_V \ll \omega_*$, while
$q ( V) \sim 2 e^*$ for $\omega_V \gg \omega*$, with the crossover
scale $\omega_*$ given by
\beq
\omega_* \sim \frac{\pi u}{L} \left[ \frac{\bar{g}_1^2 \cos^2 ( \theta ) }{\bar{g}_2^2} \frac{\Gamma
[ 1 - \frac{8}{g} ]}{ \Gamma [ 1 - \frac{2}{g} ] } \frac{ \sin \left( \frac{8 \pi}{g} \right)}{
\sin \left( \frac{2 \pi}{g} \right) } \right]^{\frac{6}{g}}
\:\:\:\: .
\label{sper7}
\eneq
\noindent
Eq.(\ref{sper7}) states that  the contributions
to $I_{\rm DC} ( V)$ and to $S ( V )$, which are $\propto \bar{g}_1^2 \cos^2 ( \theta )  $
and $\propto \bar{g}_2^2$ respectively,  are of the same order of
magnitude.

\begin{figure}
\includegraphics*[width=1.0\linewidth]{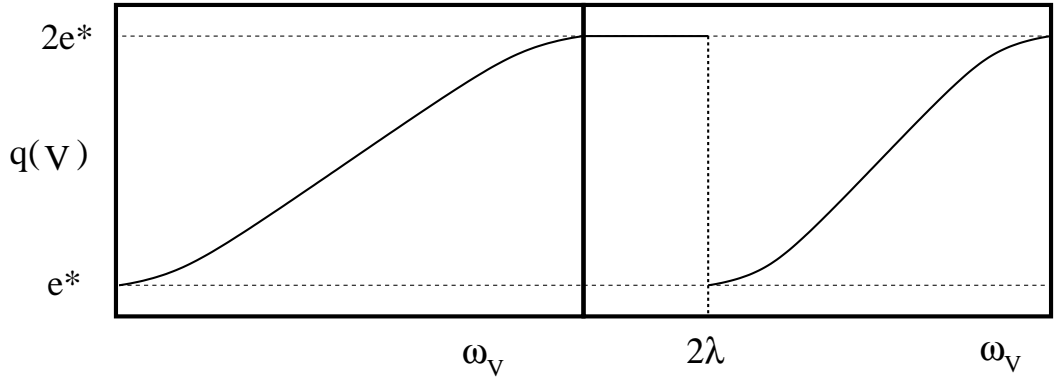}
\caption{Plot of $ q ( V ) \equiv S ( V ) / I_{\rm DC} ( V )$.
{\bf Left-hand panel:} $ q ( V )$ {\it vs.} $ \omega_V$
for $B_\parallel \neq 0$ ($\tau_1$-symmetry explicitly
broken); {\bf Right-hand panel:} $ q ( V )$ {\it vs.} $ \omega_V$
for $B_\parallel = 0$ ($\tau_1$-symmetric case). While  for
$B_\parallel \neq 0$ $q ( V)$ flows from $e^*$ (low-voltage
limit) to $2e^*$, for $B_\parallel = 0$ and $B_\perp \neq 0$
$q ( V ) = 2 e^*$ also for $ | \omega_V  | \leq
2 \lambda = 2 | B_\perp|$.}
\label{shotnoise}
\end{figure}

As $\omega_V \to 0$, the value of $q (V) $ characterizing
the IR stable fixed point is always $q = e^*$,
except when $B_\parallel = 0 , B_\perp \neq 0$, where,
the existence of  a threshold
at $\omega_V = 2 \lambda$ implies that  $q ( V) = 2 e^* $, for
$\omega_V \leq 2 \lambda = 2 | B_\perp|$.
Thus, for $B_\parallel = 0 , B_\perp \neq 0$
$4e$ superconductivity is evidenced  in the dissipative dc transport
 as well,  when $\omega_V \leq 2 \lambda$. $q ( V )$ is plotted in
Fig.\ref{shotnoise}, both for $B_\parallel \neq 0$, and for
$B_\parallel = 0 , B_\perp \neq 0$. For $B_\parallel = 0 ,
B_\perp = 0$, the threshold at $\omega_V = 2 \lambda$ moves to
$\omega_V = 0$. This is again consistent
with the picture of a $4e$-deconfined superconductor, with  $2e$
elementary charged excitations (see section  \ref{rg2}).

\subsection{Exact formula for $I_{\rm DC}  (V )$ and $S ( V )$ in
the fermionized theory for $g=2$}
\label{sno2}

In this subsection, we use the exact fermionization
of the BDSG Hamiltonian at $g=2$ (see appendix \ref{shot}
for details), to
compute $I_{\rm DC}  (V )$ and $S ( V )$ beyond perturbation theory,
for $B_\parallel =0$.

Applying a finite voltage bias $V$ to the leads amounts
to  add to the fermionic action $S^{\rm Fer}$ given in Eq.(\ref{wwc1}),
a contribution given by

\beq
S_V = \frac{e^* V}{2} \: \int \: d t \: \int_0^L \: d x \:
[ : \psi_R^\dagger ( x , t ) \psi_R ( x , t ) :  +
: \psi_L^\dagger ( x , t ) \psi_L ( x , t ) : ]
\:\:\:\: .
\label{fermi1}
\eneq
\noindent
In the fermionized theory, the current  operator
$j_H ( t )$  is given by
\beq
j_H (t) = e^* u \{ : \psi_R^\dagger ( 0 ,t  ) \psi_R ( 0 , t ) :
- : \psi_L^\dagger  ( 0 , t) \psi_L ( 0 , t ) : \}
\:\:\:\: .
\label{fermi2}
\eneq
\noindent
Taking into account that the finite voltage bias shifts
the chemical potential of the $~_R$ ($~_L$) modes
by $e^*V/2$ ($-e^*V/2$) and using  the linear relations
between the right-handed modes $\psi_R ( p )$ and the
left-handed modes $\psi_L ( p )$ provided in Eq.(\ref{wcou7}),
one obtains that the average current $I_{\rm DC} ( V ) $ is given by

\beq
I_{\rm DC}  ( V )  =  \frac{e^*u}{L} \: \sum_{ 0 \leq p \leq ( e^* V /2u)} \: [ 1 - | R ( p ) |^2 +
| Q ( p ) |^2 ] = \frac{2e^*u}{\pi} \int_0^{ ( e^* V /2u)}  \: d p \: | Q ( p ) |^2
\:\:\:\:,
\label{tran1}
\eneq
\noindent
where, to derive Eq.(\ref{tran1}), we
use  $ | Q ( p ) |^2 = 1 - | R ( p ) |^2$.

To compute $S ( V )$ through  Eq.(\ref{sshot1}),
one first needs to determine  $ S ( t , t' ) = \langle \{ I_{\rm fl} ( t ) , I_{\rm fl} ( t' ) \} \rangle_{\rm NE}$,
which,  in the fermionized theory, is given by  $S ( t - t' ) = \Sigma_1 ( t - t' )
+ \Sigma_2 ( t - t') + \Sigma_3 ( t - t' )$, with
 \[
        \Sigma_1 ( t - t' ) = - 2 \left( \frac{e^* u}{L} \right)^2 \sum_{ p q} e^{ i [ \epsilon_p -
\epsilon_q ] ( t - t' ) } [ 1 - R^* ( p ) R ( q ) + Q^* ( - p ) Q ( - q ) ]
\times
\]
\beq
\theta \left( \frac{e^*V}{2u} - p \right) \theta \left( - \frac{e^*V}{2u} + q \right)
       \:\:\:\: ,
\label{tran2}
\eneq
\noindent
\[ \Sigma_2 ( t - t' ) = \Sigma_3 ( t - t' )
 \]
\beq
= 2 \left( \frac{e^*u}{L} \right)^2 \sum_{ p q} e^{ i [ \epsilon_p -
\epsilon_q ] ( t - t' ) } | R ( p ) |^2 | Q ( q ) |^2
 \theta \left( \frac{e^*V}{2u} - p \right)\theta \left(  \frac{e^*V}{2u} + q \right)
       \:\:\:\: ,
\label{tran3}
\eneq
\noindent
and $\epsilon_p = u p $.

Taking the Fourier transform of $\Sigma ( t - t' )$ at frequency $\Omega$ and
setting $\Omega = 0$, one obtains

\[
S(V) =  S ( \Omega \to 0 , V ) = \frac{4 (e^*)^2 u}{L} \: \sum_{ -  ( e^* V /2u) \leq p \leq ( e^* V /2u)}
\: | Q ( p ) |^2 \: | R ( p ) |^2 =
\]
\beq
\frac{8 (e^*)^2 u}{\pi}  \: \int_0^{( e^* V /2u)} \: d p \: \{ | Q ( p ) |^2 - | Q ( p ) |^4 \}
\:\:\:\: .
\label{tran5}
\eneq
\noindent
As a result, from Eq.(\ref{amp2}) of appendix \ref{shot}, one
has

\beq
I_{\rm DC} ( V ) = \frac{2e^*u}{\pi} \int_0^{ ( e^* V /2u)}  \: d p \: \
\frac{ [  - \frac{g_1^2}{8 u } + u g_2 p^2 + 2 B_\perp g_2 p ]^2
 }{ ( u p + 2 B_\perp )^2  + [  - \frac{g_1^2}{8 u } + u g_2 p^2 + 2 B_\perp g_2 p ]^2 }
\:\:\:\: ,
\label{excu1}
\eneq
\noindent
and

\beq
S ( V ) = \frac{8 (e^*)^2 u}{\pi}  \: \int_0^{( e^* V /2u)} \: d p \:
\frac{( u p + 2 B_\perp )^2  [  - \frac{g_1^2}{8 u } + u g_2 p^2 + 2 B_\perp g_2 p ]^2
 }{ \{ ( u p + 2 B_\perp )^2  + [  - \frac{g_1^2}{8 u } + u g_2 p^2 + 2 B_\perp g_2 p ]^2 \}^2}
\:\:\:\: .
\label{exsh1}
\eneq
\noindent
Both integrals in Eqs.(\ref{excu1},\ref{exsh1}) may be exactly
computed, although the final expressions are not, in general,
very enlightening. However,  for $B_\perp = 0$, one may plot
$q ( V)$ {\it vs.} $V$, obtaining a result matching
the left-hand panel of Fig.\ref{shotnoise}; in particular,
one finds that

\beq
\lim_{ V \to 0 } q ( V ) = e^* \;\;\; , \;\; \lim_{ V \to \infty } q ( V ) = 2 e^*
\:\:\:\: .
\label{exch1}
\eneq
\noindent
Eq.(\ref{exch1}) shows that the crossover between
an effective elementary charge $e^*$ and an
effective charge $2 e^*$, induced by  acting
on the applied voltage bias $V$, is generally valid,
independently of the reliability of the perturbative
computation of subsection \ref{sno1}.

For $B_\perp \neq 0$, one obtains

\beq
\lim_{ V \to 0} q ( V ) = \frac{32 e^* (2 B_\perp)^2}{16 \left( \frac{g_1^2}{8u} \right)^2
+ \frac{g_1^2}{u} \left( \frac{g_2}{u} \right)^2 ( 2 | B_\perp | ) +
( 16 +  \left( \frac{g_2}{u} \right)^4 ) ( 2 B_\perp )^2 }
\:\:\:\: ,
\label{fexchsn1}
\eneq
\noindent
which, to the fourth order in $g_1 , g_2$,  yields

\beq
\lim_{ V \to 0 } q ( V ) = 2 e^*
\:\:\:\: .
\eneq
\noindent
This is again consistent with the result of Eq.(\ref{sper6}),
since it shows that  $B_\perp \neq 0$  may stabilize
an IR stable fixed point exhibiting $4e$ superconductivity.

\section{Concluding remarks}
\label{concl}

We investigated a boundary double Sine-Gordon model, where two
boundary terms, of different periodicity and scale dimensions, are
coupled to a Kondo-like spin degree of freedom.  We showed that
the pertinent engineering of the coupling between the spin degree
of freedom and the bosonic field induces a competition  between
the two boundary terms, and that this gives rise to
nonperturbative phenomena, such as the emergence of novel quantum
phases. We showed indeed that the strongly coupled fixed point- in
a pertinent range of parameters- becomes unstable as a result of
the deconfinement of new phase-slip operators (i.e., the short
instantons), arising from adding the less relevant boundary
operator.

To look for a physical context where such nonperturbative
phenomena may be observed, we analyzed  a Josephson junction
network, providing a remarkable realization of  the BDSG field
theory described by $H_B$. For this network, we showed that
the competition between the two periodicities stabilizes a robust
finite coupling fixed point and allows for the emergence, at a
pertinent scale, of $4e$ superconductivity. To probe the latter
phenomenon, we computed the dc shot noise and showed that, by
pertinently tuning the applied voltage bias $V$, the effective
charge of the carriers varies from $2e$ to $4e$.

In our analysis the onset of the  $\tau_1$ symmetry is tuned
by acting upon the control parameter $B_\parallel$. The role of
this symmetry is twofold. In section \ref{rg1} we showed that, for
$B_\perp \neq 0$, $4e$ superconductivity is naturally associated
to the realization of this symmetry. On the other hand, in section
\ref{rg2} we showed that, for $B_\perp = 0$, $\tau_1$-symmetry
implies SI deconfinement (and, thus, for $1<g<4$ it stabilizes the
IR stable FFP); at variance, short instantons are confined, if
$\tau_1$-symmetry is "broken", and the SFP is IR stable for
$1<g$.

We argued that while, as the system size $L \to \infty$, the
FFP is IR stable only when $B_\parallel = B_\perp = 0$,  for a
realistic JJN of finite size L, the FFP is robust also against
small fluctuations of the control parameters $B_\parallel ,
B_\perp$, provided that $u/L$ is sufficiently big. Indeed, as
discussed in detail in appendix \ref{instconf}, SI deconfinement
(which is the mechanism de-stabilizing the SFP {\it vs.} the FFP)
is effective only for $L \geq L_{\rm Stop} \sim 1 / (g_1^*  \cos (
\theta ) )$. As a result, for $L \leq L_{\rm Stop}$, one may
safely assume that the behavior of the device is still driven by
the FFP. This renders the emergence of the FFP in the phase
diagram relevant for engineering realistic superconducting devices
with enhanced coherence.

It may be worth to investigate how the methods developed in
this paper could be modified to account for the effects of
different commensurability ratios between competing boundary
interactions of a boundary Hamiltonian.

\newpage

\appendix

\section{Derivation of $H_{\bf B}$}
\label{swolff}

In this appendix, we describe the Schrieffer-Wolff
summation procedure, to derive the boundary interaction
Hamiltonian $H_{\bf B}$ in Eq.(\ref{el2}). In order to implement
this procedure, one should determine
the low-energy states of the central region, obtained
after diagonalizing the effective  spin-1/2 Hamiltonian

\beq
H_{\bf C} = - J \sum_{ j = 0}^3
\{ e^{ i \frac{\varphi}{4} } S_j^+ S_{j+1}^\dagger + {\rm h.c.} \} -
h \sum_{ j = 0}^3 S_j^z
\:\:\:\: .
\label{appi1}
\eneq
In Eq.(\ref{appi1}),  $h$ describes a
slight detuning of the gate voltage $V_g
 = N + 1 /2$ ($N$ integer), acting on the junctions in   ${\bf C}$.

\subsection{Eigenvalues and eigenstates of $H_{\bf C}$}

$H_{\bf C}$ commutes with $S^z = \sum_{ j =0}^3 S_j^z$; thus,
its eigenstates may be grouped in multiplets of  $S^z $.
Namely, the spectrum is given by:

\begin{itemize}

\item {\bf Spin-2 eigenstate}
There is only one state, of energy $\epsilon_2$, given by

\beq
| 2 \rangle = | \uparrow \uparrow \uparrow \uparrow \rangle
\;\;\; , \;\; \epsilon_2 = -4h
\;\;\;\; ,
\label{spectrum2}
\eneq
\noindent

\item {\bf Spin- -2 eigenstate}
There is only one state, as well, of energy $\epsilon_{-2}$,
given by

\beq
| - 2 \rangle = | \downarrow \downarrow \downarrow \downarrow \rangle
\;\;\; , \;\; \epsilon_{-2} = 4h
\;\;\;\; ,
\label{spectrum3}
\eneq
\noindent

\item {\bf Spin-1 eigenstates}
There are four spin-1 eigenstates,  with
energies given by

\beq
| 1 , k \rangle = \frac{1}{2} \sum_{ j = 0}^3 e^{ i k j } | 1 , j \rangle
\;\;\; , \;\; \epsilon_1 ( k ) = - 2 J \cos ( k - \frac{\pi}{4} ) - 2h
\:\:\:\: ,
\label{spectrum4}
\eneq
\noindent
where

\beq
k = \frac{ 2 \pi \ell}{4} \;\;\; , \;\; k = 0 , 1 , 2 ,3
\;\;\;\; .
\label{spectrum5}
\eneq
\noindent
$ | 1 , j \rangle$ is the spin-1 state where all the spins
are $\uparrow$, except the one at site $j$, which is $\downarrow$.

\item {\bf Spin- -1 eigenstates}
There are four spin-(-1) eigenstates as well,
with energies given by

\beq
| - 1 , k \rangle = \frac{1}{2} \sum_{ j = 0}^3 e^{ i k j } | - 1 , j \rangle
\;\;\; , \;\; \epsilon_{-1} ( k ) = - 2 J \cos ( k + \frac{\pi}{4} ) + 2 h
\:\:\:\: ,
\label{spectrum6}
\eneq
\noindent
where

\beq
k = \frac{ 2 \pi \ell}{4} \;\;\; , \;\; k = 0 , 1 , 2 ,3
\;\;\;\; .
\label{spectrum7}
\eneq
\noindent
$ | - 1 , j \rangle$ is the spin-1 state, where all the spins
are $\downarrow$, except the one at site $j$, which is $\uparrow$.

\item {\bf Spin-0 eigenstates}
There are six spin-0 eigenstates. They are listed below, with their
corresponding energies

{\bf State $| 0 , a \rangle$}
\[
| 0 , a \rangle = \frac{1}{2 \sqrt{2}} \{ \sqrt{2} [ | \uparrow \downarrow
\uparrow \downarrow \rangle +  | \downarrow \uparrow \downarrow \uparrow
\rangle ] - | \downarrow \downarrow \uparrow \uparrow \rangle
\]
\beq
- | \uparrow \uparrow \downarrow \downarrow \rangle -
| \uparrow \downarrow \downarrow \uparrow \rangle -
| \downarrow \uparrow \uparrow \downarrow \rangle
 \}
\;\;\; , \;\; \epsilon_a = 2 \sqrt{2} J \cos \left( \frac{\varphi}{4} \right)
\;\;\;\; ,
\label{spectrum8}
\eneq
\noindent
{\bf State $| 0 , b \rangle $}

\[
| 0 , b \rangle = \frac{1}{2 \sqrt{2}} \{ \sqrt{2} [ | \uparrow \downarrow
\uparrow \downarrow \rangle -  | \downarrow \uparrow \downarrow \uparrow
\rangle ] +i | \downarrow \downarrow \uparrow \uparrow \rangle
\]
\beq
+i  | \uparrow \uparrow \downarrow \downarrow \rangle -
i | \uparrow \downarrow \downarrow \uparrow \rangle -
i | \downarrow \uparrow \uparrow \downarrow \rangle
 \}
\;\;\; , \;\; \epsilon_b = 2 \sqrt{2} J \sin \left( \frac{\varphi}{4} \right)
\;\;\;\; ,
\label{spectrum9}
\eneq
\noindent
{\bf State $| 0 , c \rangle $}
\[
| 0 , c \rangle = \frac{1}{2 \sqrt{2}} \{  \sqrt{2} [  | \uparrow \downarrow
\uparrow \downarrow \rangle +  | \downarrow \uparrow \downarrow \uparrow
\rangle ] +  | \downarrow \downarrow \uparrow \uparrow \rangle
\]
\beq
+   | \uparrow \uparrow \downarrow \downarrow \rangle +
 | \uparrow \downarrow \downarrow \uparrow \rangle +
 | \downarrow \uparrow \uparrow \downarrow \rangle
 \}
\;\;\; , \;\; \epsilon_c = - 2 \sqrt{2} J \cos \left( \frac{\varphi}{4} \right)
\;\;\;\; ,
\label{spectrum10}
\eneq
\noindent
{\bf State $| 0 , d \rangle $}
\[
| 0 , d \rangle = \frac{1}{2 \sqrt{2}} \{ \sqrt{2} [ | \uparrow \downarrow
\uparrow \downarrow \rangle - | \downarrow \uparrow \downarrow \uparrow
\rangle ] - i | \downarrow \downarrow \uparrow \uparrow \rangle
\]
\beq
- i  | \uparrow \uparrow \downarrow \downarrow \rangle +
i | \uparrow \downarrow \downarrow \uparrow \rangle +
i | \downarrow \uparrow \uparrow \downarrow \rangle
 \}
\;\;\; , \;\; \epsilon_d = - 2 \sqrt{2} J \sin \left( \frac{\varphi}{4} \right)
\;\;\;\; ,
\label{spectrum11}
\eneq
\noindent
{\bf State $| 0 , e \rangle $}
\beq
| 0 , e \rangle = \frac{1}{ \sqrt{2}} \{ | \uparrow \uparrow \downarrow
\downarrow \rangle -  | \downarrow \downarrow \uparrow \uparrow \rangle \}
\;\;\; , \;\; \epsilon_e = 0
\;\;\;\; ,
\label{spectrum12}
\eneq
\noindent
{\bf State $| 0 , f \rangle $}
\beq
| 0 , f \rangle =  \frac{1}{ \sqrt{2}} \{ | \uparrow \downarrow \downarrow
\uparrow \rangle - | \downarrow \uparrow \uparrow \downarrow \rangle \}
\;\;\; , \;\; \epsilon_f = 0
\:\:\:\: .
\label{spectrum13}
\eneq
\noindent
\end{itemize}
For $\varphi = \pi$, the groundstate is twofold degenerate: both  states
 $ | 0 , c \rangle, | 0 , d \rangle $ have the minimum
possible energy ($-2 J$). In the next section, we consider
the effective boundary Hamiltonian arising when the
states are coupled to the leads.

\subsection{Coupling {\bf C} to the leads:
effective boundary Hamiltonian}

To determine  the effective boundary Hamiltonian $H_{\bf B}$ describing
the low-energy dynamics of {\bf C} connected to the leads, one starts
by connecting  {\bf C} to the leads by  two
junctions, of nominal strength $\lambda \ll E_J , J$, with
the Hamiltonian

\beq
H_{\bf T} = - \lambda \{ e^{ \frac{i}{\sqrt{2}} \Phi_< ( 0 ) } S_0^-
+ e^{ \frac{i}{\sqrt{2}} \Phi_> ( 0 ) } S_2^-
+ {\rm h.c.} \}
\:\:\:\: ,
\label{spectrum14}
\eneq
\noindent
where ${\bf S}_0 , {\bf S}_2$ have been defined in Fig.\ref{device}.
For $\varphi \sim \pi$, one should consider the low-energy doublet
spanned by the states $ | \Psi_\uparrow \rangle =
| 0 , c \rangle $ , $| \Psi_\downarrow \rangle = | 0 , d \rangle$,
and, in order to implement the SW summation, one needs the
matrix elements of $ | \Psi_\uparrow \rangle$
or $| \Psi_\downarrow \rangle $ with the lowest-lying excited
states of $H_{\bf C}$. The relevant matrix elements are
listed below:

\begin{itemize}

\item {\bf First set}

\[
\langle 1 , 0 | H_{\bf T} | \Psi_\uparrow \rangle = - \frac{(2 + \sqrt{2}) \lambda}{
4 \sqrt{2}} [ e^{ - \frac{i}{\sqrt{2}} \Phi_< ( 0 ) }   + e^{ - \frac{i}{\sqrt{2}} \Phi_> ( 0 ) }
 ]
\]
\beq
\langle 1 , 0 | H_{\bf T} | \Psi_\downarrow \rangle =  \frac{\lambda}{
4 } [ e^{ - \frac{i}{\sqrt{2}} \Phi_< ( 0 ) }   + e^{ - \frac{i}{\sqrt{2}} \Phi_> ( 0 ) }
 ]
\;\;\;\; ,
\label{metel1}
\eneq
\noindent

\item {\bf Second set}

\[
\langle -1 , 0 | H_{\bf T} | \Psi_\uparrow \rangle =  - \frac{(2 + \sqrt{2}) \lambda}{
4 \sqrt{2}} [ e^{  \frac{i}{\sqrt{2}} \Phi_< ( 0 ) }   + e^{  \frac{i}{\sqrt{2}} \Phi_> ( 0 ) }
 ]
\]
\beq
\langle -1 , 0 | H_{\bf T} | \Psi_\downarrow \rangle =  - \frac{\lambda}{
4 } [ e^{  \frac{i}{\sqrt{2}} \Phi_< ( 0 ) }   + e^{  \frac{i}{\sqrt{2}} \Phi_> ( 0 ) }
 ]
\;\;\;\; ,
\label{metel2}
\eneq
\noindent

\item {\bf Third set}

\[
\langle 1 , \frac{\pi}{2} |  H_{\bf T} | \Psi_\uparrow \rangle =\frac{\lambda}{
4 } [ e^{  - \frac{i}{\sqrt{2}} \Phi_< ( 0 ) }   - e^{  - \frac{i}{\sqrt{2}} \Phi_> ( 0 ) }
 ]
\]
\beq
\langle 1 , \frac{\pi}{2} |  H_{\bf T} | \Psi_\downarrow \rangle =- \frac{(2 + \sqrt{2}) \lambda}{
4 \sqrt{2}} [ e^{  - \frac{i}{\sqrt{2}} \Phi_< ( 0 ) }   -  e^{  - \frac{i}{\sqrt{2}} \Phi_> ( 0 ) }
 ]
\;\;\;\; ,
\label{metel3}
\eneq
\noindent

\item {\bf Fourth set}

\[
\langle - 1 , \frac{3 \pi}{2} |  H_{\bf T} | \Psi_\uparrow \rangle = \frac{\lambda}{
4 } [ e^{  \frac{i}{\sqrt{2}} \Phi_< ( 0 ) }   - e^{   \frac{i}{\sqrt{2}} \Phi_> ( 0 ) }
 ]
\]
\beq
\langle - 1 , \frac{3 \pi}{2} |  H_{\bf T} | \Psi_\downarrow \rangle = \frac{(2 + \sqrt{2}) \lambda}{
4 \sqrt{2}} [ e^{   \frac{i}{\sqrt{2}} \Phi_< ( 0 ) }   -  e^{   \frac{i}{\sqrt{2}} \Phi_> ( 0 ) }
 ]
\:\:\:\: .
\label{metel4}
\eneq
\end{itemize}
\noindent
Using the matrix elements given in Eqs.(\ref{metel1},\ref{metel2},\ref{metel3},\ref{metel4}§),
the SW summation yields, to second order in $\lambda$, the
following contribution to the effective boundary Hamiltonian

\[
H_{\rm Eff}^{(2)} = | \Psi_\uparrow \rangle \langle \Psi_\uparrow | \frac{\lambda^2}{32} \biggl\{
( 2 + \sqrt{2})^2 \frac{[ e^{ \frac{i}{\sqrt{2}} \Phi_< ( 0 ) } + e^{ \frac{i}{\sqrt{2}}\Phi_> ( 0 ) } ]
[ e^{ -  \frac{i}{\sqrt{2}}\Phi_< ( 0 ) } + e^{ - \frac{i}{\sqrt{2}} \Phi_> ( 0 ) } ]
}{- ( 2 - \sqrt{2})J + 2 h }
\]
\[
+  (2 + \sqrt{2})^2 \frac{[ e^{ - \frac{i}{\sqrt{2}} \Phi_< ( 0 ) } + e^{ - \frac{i}{\sqrt{2}} \Phi_> ( 0 ) } ]
[ e^{  \frac{i}{\sqrt{2}}\Phi_< ( 0 ) } + e^{ \frac{i}{\sqrt{2}} \Phi_> ( 0 ) } ]
}{- ( 2 - \sqrt{2})J - 2 h }
\]
\[
+ 2 \frac{[ e^{ \frac{i}{\sqrt{2}} \Phi_< ( 0 ) } - e^{ \frac{i}{\sqrt{2}} \Phi_> ( 0 ) } ]
[ e^{ - \frac{i}{\sqrt{2}} \Phi_< ( 0 ) } - e^{ - \frac{i}{\sqrt{2}} \Phi_> ( 0 ) } ]
}{- ( 2 - \sqrt{2})J + 2 h }
\]
\[
+
2 \frac{[ e^{ - \frac{i}{\sqrt{2}} \Phi_< ( 0 ) } - e^{ - \frac{i}{\sqrt{2}} \Phi_> ( 0 ) } ]
[ e^{  \frac{i}{\sqrt{2}} \Phi_< ( 0 ) } - e^{  \frac{i}{\sqrt{2}} \Phi_> ( 0 ) } ]
}{- ( 2 - \sqrt{2})J - 2 h }  \biggr\}
\]
\[
| \Psi_\downarrow \rangle \langle \Psi_\downarrow |\frac{\lambda^2}{32} \biggl\{
( 2 + \sqrt{2})^2 \frac{[ e^{ \frac{i}{\sqrt{2}} \Phi_< ( 0 ) } - e^{ \frac{i}{\sqrt{2}} \Phi_> ( 0 ) } ]
[ e^{ - \frac{i}{\sqrt{2}} \Phi_< ( 0 ) } - e^{ - \frac{i}{\sqrt{2}} \Phi_> ( 0 ) } ]
}{- ( 2 - \sqrt{2})J + 2 h }
\]
\[
  +  (2 + \sqrt{2})^2 \frac{
[ e^{ - \frac{i}{\sqrt{2}} \Phi_< ( 0 ) } - e^{ - \frac{i}{\sqrt{2}} \Phi_> ( 0 ) } ]
[ e^{ \frac{i}{\sqrt{2}} \Phi_< ( 0 ) } - e^{ \frac{i}{\sqrt{2}} \Phi_> ( 0 ) } ]
}{- ( 2 - \sqrt{2})J - 2 h }
\]
\[
+ 2 \frac{[ e^{ \frac{i}{\sqrt{2}} \Phi_< ( 0 ) } + e^{ \frac{i}{\sqrt{2}} \Phi_> ( 0 ) } ]
[ e^{ -  \frac{i}{\sqrt{2}}\Phi_< ( 0 ) } + e^{ -  \frac{i}{\sqrt{2}}\Phi_> ( 0 ) } ]
}{- ( 2 - \sqrt{2})J + 2 h }
\]
\[
 + 2 \frac{[ e^{ - \frac{i}{\sqrt{2}} \Phi_< ( 0 ) } + e^{ - \frac{i}{\sqrt{2}} \Phi_> ( 0 ) } ]
[ e^{   \frac{i}{\sqrt{2}}\Phi_< ( 0 ) } + e^{  \frac{i}{\sqrt{2}} \Phi_> ( 0 ) } ]
}{- ( 2 - \sqrt{2})J - 2 h }  \biggr\}
\]
\[
-  [ | \Psi_\uparrow \rangle \langle \Psi_\downarrow | +
| \Psi_\downarrow \rangle \langle \Psi_\uparrow | ]
 \frac{\sqrt{2} ( 2 + \sqrt{2} )\lambda^2}{32} \biggl\{
 \frac{[ e^{ \frac{i}{\sqrt{2}} \Phi_< ( 0 ) } + e^{ \frac{i}{\sqrt{2}} \Phi_> ( 0 ) } ]
[ e^{ - \frac{i}{\sqrt{2}} \Phi_< ( 0 ) } + e^{ - \frac{i}{\sqrt{2}} \Phi_> ( 0 ) } ]
}{- ( 2 - \sqrt{2})J + 2 h }
\]
\[
-
\frac{[ e^{ - \frac{i}{\sqrt{2}} \Phi_< ( 0 ) } + e^{ - \frac{i}{\sqrt{2}} \Phi_> ( 0 ) } ]
[ e^{  \frac{i}{\sqrt{2}} \Phi_< ( 0 ) } + e^{  \frac{i}{\sqrt{2}} \Phi_> ( 0 ) } ]
}{- ( 2 - \sqrt{2})J - 2 h }
\]
\[
+ \frac{[ e^{ \frac{i}{\sqrt{2}} \Phi_< ( 0 ) } - e^{ \frac{i}{\sqrt{2}}\Phi_> ( 0 ) } ]
[ e^{ - \frac{i}{\sqrt{2}} \Phi_< ( 0 ) } - e^{ - \frac{i}{\sqrt{2}} \Phi_> ( 0 ) } ]
}{- ( 2 - \sqrt{2})J + 2 h }
\]
\beq
 -
\frac{[ e^{ - \frac{i}{\sqrt{2}} \Phi_< ( 0 ) } - e^{ - \frac{i}{\sqrt{2}} \Phi_> ( 0 ) } ]
[ e^{   \frac{i}{\sqrt{2}}\Phi_< ( 0 ) } - e^{  \frac{i}{\sqrt{2}} \Phi_> ( 0 ) } ]
}{- ( 2 - \sqrt{2})J + 2 h } \biggr\}
\;\;\;\; .
\label{metel5}
\eneq
\noindent
Notice that, in Eq.(\ref{metel5}),  $\varphi = \pi$.
A slight detuning of $\varphi$ off $\varphi = \pi$ just
adds to $H_{\rm Eff}^{(2)}$ and effective energy splitting term,
given by

\beq
\delta H_\parallel =  4 J \sin \left( \frac{\varphi - \pi}{4} \right)
| \Psi_\uparrow \rangle \langle \Psi_\downarrow | \equiv
- B_\parallel {\bf S}^z
\:\:\:\: ,
\label{metel5.a}
\eneq
\noindent
where, in Eq.(\ref{metel5.a}), one sets $B_\parallel =
- 4 J \sin \left( \frac{\varphi - \pi}{4} \right) $ and
${\bf S}^z = | \Psi_\Uparrow \rangle \langle \Psi_\Uparrow | -
| \Psi_\Downarrow \rangle \langle \Psi_\Downarrow | $.
Expanding Eq.(\ref{metel5}) to the leading order in the
detuning parameter $h$, one finds

\beq
 H_{\rm Eff}^{(2)}  \approx - \frac{\lambda^2}{2 J }  \left(
\frac{1 + \sqrt{2}}{2 - \sqrt{2}} \right) \: {\bf S}^z \:
\cos \left[ \frac{ \Phi_> ( 0 ) -
\Phi_< ( 0 )}{\sqrt{2}} \right]
 - \frac{  \sqrt{2} ( 2 + \sqrt{2} ) }{ ( 2 - \sqrt{2} )^2 }
\: \frac{ \lambda^2 h}{J^2} {\bf S}^x
\:\:\:\: ,
\label{metel5.b}
\eneq
\noindent
with ${\bf S}^x = | \Psi_\Uparrow \rangle \langle \Psi_\Downarrow |
+ | \Psi_\Downarrow \rangle \langle \Psi_\Uparrow | $.
The BDSG Hamiltonian is recovered by  following the same procedure, up to
fourth-order in $\lambda$. The additional term is given by

\beq
H_{\rm Eff}^{(4)} =  \frac{ C \lambda^4}{  J^3}
[e^{ \frac{i}{\sqrt{2}} [ \Phi_< ( 0 ) - \Phi_> ( 0 ) ]}
 - e^{ - \frac{i}{\sqrt{2}} [ \Phi_< ( 0 )  - \Phi_> ( 0 ) ]} ]^2
+ \ldots
\;\;\;\; ,
\label{5term}
\eneq
\noindent
where $C \sim 10^{-1}$ is a numerical coefficient and
the ellipses stand for subleading contributions.
Defining $\Phi = ( \Phi_< - \Phi_> ) / \sqrt{2}$,
$g_1 = \frac{\lambda^2}{2 J }  \left(
\frac{1 + \sqrt{2}}{2 - \sqrt{2}} \right) $, $g_2 =
2 \frac{ C \lambda^4}{  J^3} $ and $B_\perp =
\frac{  \sqrt{2} ( 2 + \sqrt{2} ) \lambda^2}{ ( 2 - \sqrt{2} )^2 }
\frac{ \lambda^2 h}{J^2} $, one obtains the effective boundary
Hamiltonian given in Eq.(\ref{el2}), as
$H_{\bf B} = H_{\rm Eff}^{(2)} + H_{\rm Eff}^{(4)} +
\delta H_\parallel $. We notice that both $B_\perp$ and $B_\parallel$ may
be tuned by acting on external control fields: $B_\perp$ with  $V_g$,
$B_\parallel$  with  $\varphi$.

\section{Instanton solutions of the boundary double Sine-Gordon model and short instanton
deconfinement}
\label{instconf}

In this appendix, we study in detail  instanton solutions of the
BDSG model, regarded  as imaginary time trajectories of  the zero mode, $P ( \tau )$.
To construct the effective Euclidean action for  $P ( \tau )$, $S_E [ P ] $, one
uses for $\Phi ( x , \tau )$ the mode expansion in Eq.(\ref{sc1}).
$S_E [ P ] $ is computed from

\beq
e^{ - S_E [ P ] } = \int {\bf D} \Phi_{\rm osc}  {\bf D} \{a , b \} \: e^{ - S_E^{(0)} [ \Phi ] -
S_{\bf B} }
\:\:\:\: ,
\label{appisc1}
\eneq
\noindent
where   $\int {\bf D} \Phi_{\rm osc}  {\bf D} \{a , b \} \:$ denotes
functional integration over the oscillator modes of the field $\Phi$,
as well as over the local fermionic operators, associated with
the spin ${\bf S}$. The free
Euclidean action for the field $\Phi ( x , \tau )$ is given by

\beq
S_E^{(0)}[ \Phi ]  = \frac{g}{4 \pi}\: \int_0^\beta \: d \tau \: \int_0^L \: d x \:
\left[ \frac{1}{u} \left( \frac{\partial \Phi}{\partial \tau} \right)^2 +
u \left( \frac{\partial \Phi}{\partial x} \right)^2 \right]
\:\:\:\: .
\label{appe1}
\eneq
\noindent
From the  the mode expansion given in Eq.(\ref{sc1}),
one obtains

 \beq
 S_E [ P ] = S_E^{(0)} [ P ] + \delta S_E [ P ]
\:\:\:\: ,
\label{appe1.b}
\eneq
\noindent
with
\beq
S_E^{(0)} [ P ]  = \int_0^\beta \:  d \tau  \left\{ \frac{M}{2} \:
( \dot{P} )^2 \:+
\frac{ \pi}{ u L} P^2 - \bar{g}_1  \cos ( \theta ) \cos \left[ \frac{\pi}{\sqrt{g}}
 P - \alpha \right] -
\bar{g}_2 \cos \left[ \frac{2\pi}{\sqrt{g}} P - 2 \alpha \right] \right\}
\:\:\:\: ,
\label{appe3}
\eneq
\noindent
$M =  \pi L / 6 u$, and $\delta S_E [ P ] $ defined by

\beq
e^{ - \delta S_E [ P ] } = \int \: {\bf D}  \Phi_{\rm osc} \:
e^{ -  S_E^{(0)} [ \Phi ] }
\:\:\:\: .
\label{appe4}
\eneq
\noindent
From Eq.(\ref{appe3}), one sees that the inductance energy ($\propto L^{-1}$)
breaks, in the finite-size system,  the degeneracy between
the minima  of $H_{\bf B}$; however,  a  degeneracy between
only nearest neighboring minima of $H_{\bf B}$ may be
restored by setting $\alpha = \pi + 2 k \pi$, if $B_\parallel > 0$,
or $\alpha = 2 \pi k$, if $B_\parallel < 0$.

Assuming  $B_\parallel > 0$ and $\alpha = - \pi$,
a single-instanton solution,  representing a quantum
jump between $p_0$ and $p_{-1}$  may be built by requiring that
 $\frac{\delta S_E^{(0)} [ P]  }{\delta P }= 0$.
Defining $p ( \tau ) = ( P ( \tau ) - 1/2 )/ \sqrt{g}$,
one obtains the following  imaginary time ``equation of motion''
for $p ( \tau )$:

\beq
M \ddot{p} - \frac{2 g \pi}{ u L } ( p - \frac{1}{2} )  +
\pi \bar{g}_1  \cos ( \theta ) \cos [ \pi p ] + 2 \pi
\bar{g}_2 \cos [ 2 \pi p] = 0
\;\;\;\; .
\label{appe5}
\eneq
\noindent
Apart from the term $\propto \frac{1}{L}$, Eq.(\ref{appe5})
 is the imaginary-time version of the
equation yielding static, finite energy, soliton solutions in the
double Sine-Gordon model \cite{dsg2}. Borrowing well-known results
\cite{dsg2}, one may write down a
single-instanton solution as

\beq
p ( \tau ) = - 1 + \frac{2}{\pi} \left\{ {\rm arctan}
\left[ \exp \left( \frac{ \tau + R    }{
{\bf T}  } \right) \right] -{\rm arctan}
\left[ \exp \left( \frac{ - \tau + R   }{
{\bf T}  }\right) \right] \right\}
\;\;\;\; .
\label{appe6}
\eneq
\noindent
In Eq.(\ref{appe6}), the ``bare'' parameter $R$ is defined by the condition
$\frac{1}{4} \sinh^2 \left[ \left( \bar{g}_1 \cos ( \theta )  +
\frac{\bar{g}_2}{4}\right)R \right] =
\frac{ \bar{g}_2 }{ \bar{g}_1 \cos ( \theta )  }$. In our analysis,
 due to the logarithmic  divergences induced by interaction with the oscillator modes,
a logarithmic term in $R$ is induced. Due to this,
we regard $R$ as a variational parameter and use  the ``instanton size''
 ${\bf T}$, as the main scaling parameter.

From Eq.(\ref{appe6}),  $p ( \tau )$ may be represented as a
sequence of two short instantons, separated by a distance $2R$ from each other.
In addition,  Eq.(\ref{appe6}) allows to compute also the Euclidean action for
$p ( \tau )$, yielding
\beq
S_0 \approx - g \ln \left( \frac{ u {\bf T}}{L} \right) - g
\ln \left( \frac{ u R}{L} \right) -
\frac{4 g R}{\pi u L} + 2 \bar{g}_1  \cos ( \theta ) R
\:\:\:\: ,
\label{appe7}
\eneq
\noindent
from which, one gets
\beq
e^{ - S_0} \propto  R^g \: \exp \left[ \frac{4 g R}{\pi u L} - 2 \bar{g}_1 \cos ( \theta )  R  \right]
\:\:\:\: .
\label{appe7.1}
\eneq
\noindent

Eq.(\ref{appe7.1}) implies that
large $R$-solutions may either be favored, or disfavored, according to whether
 $\frac{2 g R}{\pi u L \bar{g}_1} > 1$, or
 $\frac{2 g R}{\pi u L \bar{g}_1} < 1$.
Since $\frac{2 g R}{\pi u L \bar{g}_1} \sim L^{\frac{1}{g} - 1}$,
for $g>1$, one sees that solutions with the two SIs at large separations are  strongly suppressed,
as $L$ goes large. The optimal value of the variational parameter  $R$ is
 set by requiring that   $\frac{d S_0}{ d R} = 0$, which, for
for small $\bar{g}_1 \cos ( \theta )$, leads to  $R_*
\propto 1 / g_1 \cos ( \theta )$.

\begin{figure}
\includegraphics*[width=.7\linewidth]{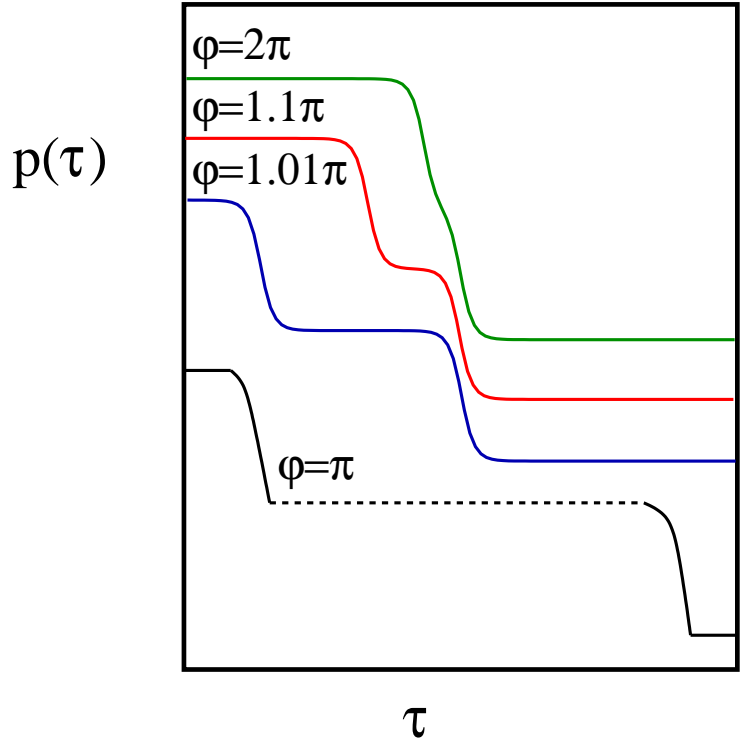}
\caption{Instanton profile $p ( \tau )$ for
various values of $\varphi$. From top to bottom: $\varphi = 2 \pi , 1.1\pi ,
1.01\pi , \pi$. The plots have been vertically shifted: in any
case $ p ( \tau )$ ranges from 0 to 1. The dashed portion of
the plot for $\varphi = \pi$ (black line) denotes infinite
separation between the short instantons ($R \to \infty$).}
\label{shdec}
\end{figure}

One should notice that,
since $R$ is directly related to  $\cos ( \theta ) g_1$,
one may readily tune it by just acting on  the applied flux $\varphi$.
Since SIs are confined over a scale $L_{\rm Stop} \sim R_*$, the
scaling of the parameter $\Lambda_1 ( L ) $ in Eq.(\ref{sc3.a}) will stop
at $L = L_{\rm Stop}$.  To show how acting on $\varphi$
may trigger SI dconfinement, in Fig.\ref{shdec} we plot
the solution $p ( \tau )$ for different values of $\varphi$.
The SI deconfinement may clearly be seen, for $\varphi = \pi$,

Finally, we mention that
the solution for $B_\parallel < 0$ and $\alpha = 0$, has
the same form as the one in Eq.(\ref{appe6}), provided that one substitutes $B_\parallel$ with
$ | B_\parallel | $.

\section{Tables of relevant integrals}
\label{integr}

\begin{figure}
\includegraphics*[width=.55\linewidth]{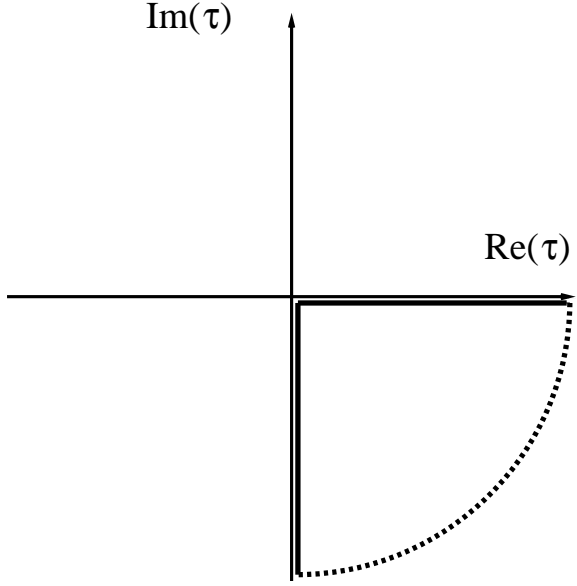}
\caption{Integration path used to compute $I_\zeta  ( z )$ in
Eq.(\ref{iden1})}
\label{intpath}
\end{figure}

In this appendix, we sketch the calculation of the integrals used in
subsection \ref{sno1} to compute $I_{\rm DC} ( V )$ and
$S ( V )$ near by the WFP. All the relevant integrals may be
recasted in the form

\beq
I_\zeta  ( z ) = \int_0^\infty \: d \tau \: \frac{e^{ - i z \tau} }{[ e^{ \frac{i \pi u}{L} \tau}
- e^{  -\frac{i \pi u}{L} ( \tau - i \eta) } ]^\zeta}
\:\:\:\: ,
\label{iden1}
\eneq
\noindent
with $\zeta$ real and $\eta = 0^+$. To compute the integral, one
 assumes $z - \zeta \pi u/L < 0$, and integrates
over the closed integration path shown in Fig.\ref{intpath}, to
get

\beq
I_\zeta   ( z ) = - i \int_0^\infty \: d w \: \frac{
e^{ - \left( \frac{\pi u \alpha}{L} + z\right)w}}{[ e^{ \frac{\pi u \eta}{L} }
- e^{ - \frac{2 \pi w }{L} }]^\zeta}
\:\:\:\: .
\label{iden2}
\eneq
\noindent

The integral in Eq.(\ref{iden2}) is tabulated \cite{astegun},
yielding

\beq
I_\zeta   ( z ) = - i  ~_2F_1 [  \frac{L z}{2 \pi u } +
            \frac{\zeta}{2} , \zeta , 1 + \frac{L z}{2 \pi u } +
            \frac{\zeta}{2} ; 1 ] / [ z - \frac{\zeta \pi u }{L}]
\:\:\:\: ,
\label{iden3}
\eneq
\noindent
where  $~_2F_1 $ is the confluent hypergeometric function.
Then, using the identity

\beq
~_2F_1 [ a , b , c ; 1 ] = \frac{\Gamma [ c ] \Gamma [ c - a - b]}{\Gamma [ c-a]
\Gamma [ c-b]}
\:\:\:\: ,
\label{iden5}
\eneq
\noindent
in Eq.(\ref{iden3}), one obtains Eqs.(\ref{sper3},\ref{sper4}) of
subsection \ref{sno1}. To analytically extend Eq.(\ref{iden3}),
one may use Stirling's approximation, which yields

\beq
\Gamma [ w ] \approx \sqrt{2 \pi} \: ( w - 1)^{ w - \frac{1}{2}} \:
e^{ - ( w - 1 ) }
\:\:\:\: .
\label{iden6}
\eneq
\noindent
Eq.(\ref{iden6}) has been applied in subsection \ref{sno1}, when
extracting the large-$L$ limit of $I_{\rm DC} ( V )$ and of  $S ( V ) $.

\section{Fermionization and exact solution of the BDSG model
for $g=2$ }
\label{shot}

In this appendix we carry out the fermionization procedure of
 $H_{\rm BDSG} = H_{\rm LL} + H_{\bf B}$ for $g=2$. As
we will show, introducing a complete set of fermionic coordinates
to represent the relevant fields of the model, allows for
recasting  $H_{\rm BDSG} $ in a quadratic form, thus
making it exactly solvable. In the following, we assume $B_\parallel =0$
and, due to our interest in  nonequilibrium dc transport properties,
we resort to the real time formalism.

The fermionization of $H_{\rm BDSG} $ follows the same
basic steps as in Ref.\cite{ameduri}. To begin with, one
fermionizes $ H_{\rm LL} $ for $g=2$ by writing
$\Phi$ as the sum of two chiral fields, $\phi_R , \phi_L$, as $\Phi ( x , t ) =
\phi_R ( x - u t ) + \phi_L ( x + u t )$. The
chiral vertex operators $: e^{ - i \phi_R ( x - u t ) }:$,
$:e^{  - i \phi_L ( x + u t ) }:$ may be the regarded as two
chiral fermionic fields $\psi_R , \psi_L$

\beq
\psi_R ( x - u t ) = \eta_R \frac{1}{\sqrt{L}} \: : e^{ - i \phi_R ( x - u t ) }: \;\;\; , \;\;
\psi_L ( x + u t ) = \eta_L \frac{1}{\sqrt{L}} \: : e^{  - i \phi_L ( x + u t ) }:
\:\:\:\: ,
\label{wcou1}
\eneq
\noindent
where $\eta_R , \eta_L$ are the Klein factors,
introduced to ensure the correct anticommutation
relations between $\psi_R$ and $\psi_L$  \cite{shulz}.

To fermionize $H_{\bf B}$, one
assumes  Neumann BCs at both
boundaries \cite{ameduri}; namely
\beq
\psi_R ( 0 , t ) = \psi_L ( 0 , t ) \;\;\;  ,\;\;
\psi_R ( L , t ) = \psi_L ( L , t )
\:\:\:\: .
\label{wcou2}
\eneq
\noindent
Eq.(\ref{wcou2}) is enforced if one uses \cite{ameduri} the noninteracting
 real time  action for $\psi_R , \psi_L$,
\[
S_0^{\rm Fer} = i \: \int \: d t \: \int_0^L \: d x \: \left\{ \psi_R^\dagger ( 0 , t )
\left[ \frac{\partial }{\partial t} + u \frac{\partial }{\partial x} \right]
\psi_R ( 0 , t ) + \psi_L^\dagger ( 0 , t )
\left[ \frac{\partial }{\partial t} - u \frac{\partial }{\partial x} \right]
\psi_L ( 0 , t )  \right\}
\]
\beq
+ i \frac{u}{2} \: \int \: d t \: \left\{ \psi_R^\dagger ( 0 , t ) \psi_L ( 0 , t ) -
\psi_L^\dagger ( 0 , t ) \psi_R ( 0 , t ) \right\}
\:\:\:\: .
\label{wcou3}
\eneq
\noindent
Then, one has to follow a different procedure for the term
$\propto g_1$ and the term $\propto g_2$ in $H_{\bf B}$.
Fermionizing the former, requires introducing
additional local degrees of freedom $\xi , \xi^\dagger$, describing the
spin-1/2 variable ${\bf S}$ emerging at {\bf C}.
Using a ``rotated'' Jordan-Wigner (JW) transformation \cite{shulz},
one introduces a complex fermion $\xi$, in terms of which
${\bf S}^x , {\bf S}^z$ are given by

\beq
{\bf S}^x \to  \xi^\dagger \xi - \frac{1}{2} \;\;\; , \;\;
{\bf S}^z \to \frac{\xi + \xi^\dagger}{2}
\:\:\:\: .
\label{wcou3.a}
\eneq
\noindent
Using Eqs.(\ref{wcou1},\ref{wcou3.a}), and
taking into account the boundary conditions in Eq.(\ref{wcou2}),
one gets the contribution to boundary action which is $\propto g_1$;
namely \footnote{Notice that $g_1$, not $\bar{g}_1$, appears in Eq.(\ref{wcou4}),
as the scale dependent factor has been reabsorbed in the
definition of the fermionic fields, Eq.(\ref{wcou2})}.

\beq
S_B^{(I)} = i \frac{g_1}{4} \: \int \: d t \: \{ [ \psi_R^\dagger ( 0 , t ) +
\psi_L ( 0 , t ) ] \xi - \xi^\dagger [ \psi_R ( 0 , t ) + \psi_L^\dagger  ( 0 , t )] \}
+ i \int \: d t \: \xi^\dagger \frac{\partial \xi ( t )}{ \partial t}
\:\:\:\: .
\label{wcou4}
\eneq
\noindent
To fermionize the contribution to the boundary interaction which is
proportional to $g_2$, one has to regularize, by point splitting, the products
$: e^{ a i \phi_{A } ( 0 )} : \: : e^{ b i \phi_{B } ( 0 )} :$,
with $a , b = \pm 1$ and $A , B = L , R$, and to require that
the anticommutation relations between the
chiral fermions are preserved. As a result, one gets
\[
S_B^{(II)} =  i \frac{ g_2}{2}  \int \: d t \:  \left\{
\psi_R^\dagger ( 0 , t ) \frac{ \partial \psi_L^\dagger ( 0 , t ) }{
\partial t}  + \psi_L ( 0 , t ) \frac{ \partial \psi_R ( 0 , t ) }{
\partial t}  \right\}
\]
\beq
i \frac{ g_2 }{2}  \int \: d t \:  \left\{
 \psi_R^\dagger ( 0 , t ) \frac{ \partial \psi_R ( 0 , t  ) }{
\partial t} + \psi_L^\dagger ( 0 , t ) \frac{ \partial \psi_L ( 0 , t ) }{
\partial t} \right\}
\:\:\:\: .
\label{wcou8}
\eneq
\noindent
Finally, the term $\propto B_\perp$ is
fermionized using the JW transformation
reported in Eq.(\ref{wcou3.a}) and, by adding the ``kinetic''
term for $\xi$, one gets the
last contribution to the boundary action, which is given by

\beq
 S_B^{(III)} =\int \: d t \:\left[  i \xi^\dagger \dot{\xi} + 2 B_\perp \left( \xi^\dagger
\xi - \frac{1}{2} \right) \right]
\:\:\:\: .
\label{wcou8.a}
\eneq
From Eqs.(\ref{wcou3},\ref{wcou4},\ref{wcou8},\ref{wcou8.a}), one
sees that the full fermionic action is given by

\beq
S^{\rm Fer} = S_0^{\rm Fer} + S_B^{(I)} + S_B^{(II)}+ S_B^{(III)}
\:\:\:\: .
\label{wwc1}
\eneq
\noindent
Equating to zero the functional derivative of $S^{\rm Fer}$,
one obtains the boundary conditions for the fermionic fields, which
are  given by

\[
 - i \frac{u}{2} [ \psi_R ( 0 , t )  - \psi_L ( 0 , t ) ] + i \frac{g_2}{2} \left(
\frac{ \partial \psi_L^\dagger ( 0 , t ) }{ \partial t} +
\frac{ \partial \psi_R ( 0 , t ) }{ \partial t}  \right) + i \frac{g_1}{2} \xi = 0
\]
\beq
i \dot{\xi} + B_\perp \xi - i \frac{g_1}{4} [ \psi_R ( 0 , t ) + \psi_L^\dagger ( 0 , t ) ] = 0
\:\:\:\: .
\label{wcou9}
\eneq
\noindent
Getting rid of $\xi , \dot{\xi}$ in Eqs.(\ref{wcou9}) and using the normal mode
expansion for $\psi_R , \psi_L$, given by

\beq
\psi_R ( x - u t ) = \frac{1}{ \sqrt{L}} \: \sum_p e^{ i ( x - u t )} \: \psi_R ( p ) \;\;\; ,
\;\; \psi_L( x + u t ) = \frac{1}{ \sqrt{L}} \: \sum_p e^{ i ( x + u t )} \: \psi_L ( p ) \;\;\;\;
,
\label{wcou10}
\eneq
\noindent
one eventually obtains the following linear relations between the
normal modes $\psi_R ( p ) , \psi_L ( p )$

\begin{eqnarray}
 \psi_R (  p ) &=& R ( p ) \psi_L ( - p )  + Q ( p ) \psi_L^\dagger (  p ) \nonumber \\
 \psi_R^\dagger ( - p ) &=& Q ( p ) \psi_L ( -  p )  + R ( p ) \psi_L^\dagger ( p )
\:\:\:\: ,
\label{wcou7}
\end{eqnarray}
\noindent
with
\beq
R (p) = \frac{ u p + 2 B_\perp }{ u p + 2 B_\perp  - i [  - \frac{g_1^2}{8 u } + u g_2 p^2 + 2 B_\perp g_2 p ]}
\;\;\;\; ,
\label{amp1}
\eneq
\noindent
and
\beq
Q (p) = \frac{i [  - \frac{g_1^2}{8 u } + u g_2 p^2 + 2 B_\perp g_2 p ]
 }{ u p + 2 B_\perp  - i [  - \frac{g_1^2}{8 u } + u g_2 p^2 + 2 B_\perp g_2 p ] }
\;\;\;\; .
\label{amp2}
\eneq
\noindent
Eqs.(\ref{amp1},\ref{amp2}) provide the formulas we
used in subsection \ref{sno2} to compute $I_{\rm DC} ( V)$
and $S ( V )$ in the fermionized theory.

\end{document}